\documentclass[%
 reprint,
 amsmath, amssymb,
 aps,
]{revtex4-2}
\usepackage{graphicx}
\usepackage{dcolumn}
\usepackage{bm}

\newcommand{\dd}{\mathrm{d}}
\newcommand{\DD}{\mathcal{D}}

\newcommand{\lrp}[1]{\left( #1 \right)}
\newcommand{\lrs}[1]{\left[ #1 \right]}
\newcommand{\lrc}[1]{\left\{ #1 \right\}}

\newcommand{\IB}{\text{IB}}
\DeclareMathOperator{\argmin}{argmin}

\newcommand{\RN}[1]{%
\textup{\uppercase\expandafter{\romannumeral#1}}%
}

\begin{document}

\preprint{APS/123-QED}

\title{Gaussian Information Bottleneck and the Non-Perturbative Renormalization Group}

\author{Adam G. Kline}
\affiliation{%
 Department of Physics, The University of Chicago, Chicago IL 60637
}
\author{Stephanie E. Palmer}%
\affiliation{%
 Department of Organismal Biology and Anatomy and Department of Physics, The University of Chicago, Chicago IL 60637
}%

\date{\today}

\begin{abstract}

The renormalization group (RG) is a class of theoretical techniques used to explain the collective physics of interacting, many-body systems. It has been suggested that the RG formalism may be useful in finding and interpreting emergent low-dimensional structure in complex systems outside of the traditional physics context, such as in biology or computer science. In such contexts, one common dimensionality-reduction framework already in use is information bottleneck (IB), in which the goal is to compress an ``input'' signal $X$ while maximizing its mutual information with some stochastic ``relevance'' variable $Y$. IB has been applied in the vertebrate and invertebrate processing systems to characterize optimal encoding of the future motion of the external world. Other recent work has shown that the RG scheme for the dimer model could be ``discovered'' by a neural network attempting to solve an IB-like problem. This manuscript explores whether IB and any existing formulation of RG are formally equivalent. A class of soft-cutoff non-perturbative RG techniques are defined by families of non-deterministic coarsening maps, and hence can be formally mapped onto IB, and vice versa. For concreteness, this discussion is limited entirely to Gaussian statistics (GIB), for which IB has exact, closed-form solutions. Under this constraint, GIB has a semigroup structure, in which successive transformations remain IB-optimal. Further, the RG cutoff scheme associated with GIB can be identified. Our results suggest that IB can be used to \textit{impose} a notion of ``large scale'' structure, such as biological function, on an RG procedure.

\end{abstract}

\maketitle


\section{Introduction}

An overarching theme in the study of complex systems is effective low-dimensionality. We are content, for example, with the existence of laws of fluid dynamics whose few phenomenological parameters accurately account for the macroscopic behavior of many completely different fluids. We are also confident that the laws are insensitive to the particular microscopic configuration of a fluid at any given time. These are  connected, but different notions of low-dimensionality; the first deals with simplification in model space, while the second refers to the emergence of collective modes, of which relatively few, when compared to the total number of degrees of freedom, will be important. A central result of Wilson’s renormalization group (RG) formulation is that an effective low-dimensional model of a system may be found through repeated coarsening of the microscopic or “bare” model. In other terms, by successively removing dynamical degrees of freedom from the system description, the effective model “flows” towards a description involving very few parameters. In general, there are many strategies which can be used to simplify the description of a high-dimensional system, and RG methods, though vast in breadth, form only a subset of these. An altogether different dimensionality reduction framework is the information bottleneck (IB), which attempts to compress (or more accurately \textit{coarsen}) a signal while keeping as much information about some \textit{a priori} defined ``relevance'' variable as possible \cite{tishby_information_2000}. Both IB and RG have been applied in theoretical neuroscience \cite{creutzig_past-future_2009,chalk_toward_2018,palmer_predictive_2015,meshulam_coarse_2019}, computer science \cite{mehta_exact_2014,tishby_deep_2015,shwartz-ziv_opening_2017,alemi_deep_2019,kolchinsky_nonlinear_2019,saxe_information_2019}, and other frontier areas of applied statistical physics \cite{koch-janusz_mutual_2018,gordon_relevance_2021}. Given the ubiquitous need to find simplifying structure in complex models and data, a synthesis of the ideas present in IB and RG could yield powerful new analysis methods and theoretical insight.

Probability-theoretic investigations of renormalization group methods are not a recent development \cite{jona-lasinio_renormalization_2001}. One early paper by Jona-Lisinio used limit theorems from probability theory to argue the equivalence of older, field-theoretic RG formalism due to Gell-Mann and Low with the modern view due to Kadanoff and Wilson \cite{jona-lasinio_renormalization_1975}. Recent work \cite{apenko_information_2012,machta_parameter_2013,beny_renormalization_2015,raju_information_2018,lenggenhager_optimal_2020,gordon_relevance_2021} has focused on connections of RG to information theory. Since the general goal in RG is to remove information about some modes or system states through coarsening, an effective characterization of RG explains how the information loss due to coarsening generates the RG flow or relates to existing notions of emergence. Moreover, like the probabilistic viewpoint promoted by Jona-Lisinio, the information-theoretic viewpoint enjoys a healthy separation from physical context. The hope is that, by removing assumptions about the particular organization or interpretation of the degrees of freedom in the system, RG methods can be generalized and made applicable to problems outside of a traditional physics setting \cite{meshulam_coarse_2019, bradde_pca_2017}. This viewpoint also has the potential to enrich traditional RG applications, as Koch-Janusz et al. point out \cite{koch-janusz_mutual_2018}. Their neural-network implementation of an IB-like coarsening scheme was able to ``discover'' the relevant, large-scale modes of the dimer model, whose thermodynamics are completely entropic, and whose collective modes do not resemble the initial degrees of freedom. More recently, Gordon et al.\ built upon this scheme to formally connect notions of ``relevance" between IB and RG \cite{gordon_relevance_2021}.

In contrast to most RG formulations which require an explicit, \textit{a priori} notion of how the modes of the system should be ordered, the information bottleneck approach defines the relevance of a feature by the information it carries about a specified relevance variable. To be concrete, let $X$ be a random variable, called the “input,” which we wish to coarsen. Then, let $Y$ be another random variable, called the “relevance variable,” which has some statistical interaction with the input $X$. IB defines a non-deterministically coarsened version of $X$, $\tilde{X}$, which is optimal in the sense that the mutual information (MI) between $\tilde{X}$ and $Y$ is maximized.
Because $\tilde{X}$ is defined as a non-deterministic coarsening of $X$, an exact correspondence between RG and IB demands that the RG scheme uses what is known as a ``soft'' cutoff. This means, for example, that the ubiquitous perturbative momentum shell approach put forth by Wilson cannot be mapped exactly onto IB under the interpretation of $\tilde{X}$ as some coarse-grained variable.
The trade-off between degree of coarsening, indicated by $I(\tilde{X};X)$ and the amount of relevant information retained $I(\tilde{X};Y)$ is controlled by a continuous variable, denoted $\beta$. Formally, the non-deterministic map which yields $\tilde{X}$ from $X$ is found by optimizing the IB objective function:
\begin{equation}\label{eq:IB_objective}
	P_\beta(\tilde{x}|x) = \argmin_{P(\tilde{x}|x)} I(X;\tilde{X}) - \beta I(\tilde{X};Y)
\end{equation}

For large values of $\beta$, the compressed representation $\tilde{x}$ is more detailed and retains a greater deal of predictive information about $Y$. Conversely, for smaller $\beta$, relatively few features are kept, in favor of reducing $I(\tilde{X};X)$ (increasing compression/coarsening). The formalism investigated here is the one originally laid out in 2000 by Tisbhy et al. \cite{tishby_information_2000}, but since then a number of thematically similar IB schemes have been proposed \cite{slonim_multivariate_2006,slonim_agglomerative_2000,strouse_deterministic_2017}. IB methods have been employed extensively in computer science, specifically towards artificial neural networks and machine learning \cite{mehta_exact_2014,tishby_deep_2015,shwartz-ziv_opening_2017,alemi_deep_2019,kolchinsky_nonlinear_2019,saxe_information_2019}. In theoretical neuroscience, Palmer et al.\ have demonstrated using IB that the retina optimally encodes the future state of some time-correlated stimuli, suggesting that prediction is a biological function instantiated early on in the visual stream \cite{palmer_predictive_2015,salisbury_optimal_2016}. IB has also been applied in studies of other complex systems, for instance to efficiently discover important reaction coordinates in large MD simulations \cite{wang_pastfuture_2019}, and to rigorously demonstrate hierarchical structure in the behavior of Drosophila over long timescales \cite{berman_predictability_2016}.

From a broad perspective, there are some basic similarities between RG and IB. Both frameworks entail a coarsening procedure by which the irrelevant aspects of a system description are discarded in order to generate a lower-dimensional, “effective” picture. Further, the Lagrange multiplier $\beta$ in IB, which parameterizes the level of detail retained, can be seen as roughly analogous to the scale cutoff present in some implementations of RG. As a first guess, one might imagine that $X$ in IB roughly corresponds to the (fluctuating) bare state of a system we are interested in renormalizing, and its compressed representation $\tilde{X}$ is a coarsened dynamical field akin to a fluctuating “local” order parameter. However, it is not difficult to find implementations of RG which do not map to IB in this way, and vice versa. For example, in Wilsonian RG schemes with a hard momentum cutoff, the decimation step represents a deterministic map from bare to coarsened system state. Together with our provisional interpretation, this contradicts the original formulation of IB, in which the coarsening is non-deterministic \footnote{Alternative IB frameworks have been proposed which result in deterministic mappings \cite{slonim_agglomerative_2000, strouse_deterministic_2017}, and these could conceivably be connected to hard-cutoff RG schemes, though some issues occur for continuous random variables. We restrict our present discussion to the original framework.}. 

Another, more serious discrepancy is due to the expected use cases of these two theoretical frameworks. Generically, the fixed point description of criticality offered by RG is legitimate due to the presence of infinitely many interacting degrees of freedom, otherwise the coarsened model cannot be mapped back into the original model space. In IB, the random variables $X$ is finite-dimensional, such as a finite lattice of continuous spins, and “dimensional reduction” does not refer to the convergence towards a low-dimensional critical manifold in model space, but instead the actual removal of dimensions from the coarsened representation of $X$. Finally, and perhaps most dauntingly, there is not an obvious equivalent of the IB relevance variable $Y$ in RG. It seems counterintuitive that one would want more control over the collective mode basis used to describe a system, when for the vast majority of RG applications, length or energy scale works perfectly well as a cutoff. 

Despite these apparent mismatches, there are some significant structural similarities between IB for continuous variables and a class of RG implementations involving soft cutoffs. For concreteness, we restrict our discussion of the correspondence to Gaussian statistics. While this precludes the analysis of non-Gaussian criticality, it allows all of the results to be expressed analytically and makes connections more transparently. This can also serve as a basis for later investigations involving non-Gaussian statistics and interacting systems. To begin, we show that Gaussian information bottleneck (GIB) \cite{chechik_information_2005} exhibits a semi-group structure in which successive IB coarsenings compose larger IB coarsenings. This structure is summarized in an explicit function of the Lagrange multiplier $\beta$ which simply multiplies under semigroup action and is therefore analogous to the length scale in canonical RG. Next we explore how the coarsening map $P(\tilde{x}|x)$ provided by IB defines an infra-red regulator which serves as a soft cutoff in several non-perturbative renormalization group (NPRG) schemes. This relation shows that the freedom inherent in choosing a cutoff scheme maps directly to the choice of $Y$-statistics in IB. Finally, we use a Gaussian field theory as a toy model to explore the physical significance of this fact. One result is that the RG scheme provided by IB can select a collective mode basis which is not Fourier, and hence impose a cutoff which cannot be interpreted as a wavenumber. Additionally, in whichever collective mode basis is chosen, the shape of this IB cutoff scheme is closely related to the Litim regulator which is ubiquitous in NPRG literature \cite{litim_optimized_2001}.

\section{Semigroup structure in Gaussian Information Bottleneck \label{sec:semigroup}} 

Every IB problem begins with the distribution $P(x,y)$, which specifies the statistical dependencies linking the input variable $X$ to the relevance variable $Y$.
Gaussian information bottleneck (GIB) refers to the subset of IB problems in which $P(x,y)$ is jointly Gaussian. Under this constraint, a family of coarsening maps $P_\beta(\tilde{x}|x)$ can be found exactly for all $\beta$. Chechik et al.\ \cite{chechik_information_2005} showed this by explicitly parameterizing the coarsening map, then minimizing the IB objective function with respect to these parameters. Their parameterization consists of two matrices $A$ and $\Sigma_\xi$, which are used to define the compressed representation $\tilde{X}$ as a linear projection of the input plus a Gaussian ``noise" variable $\xi$. Explicitly, $\tilde{X} = AX + \xi$ with $\xi \sim \mathcal{N}(0,\Sigma_\xi)$. Under this parameterization, one exact solution is given by:
\begin{equation}\label{eq:GIB_sol}
    \begin{cases}
    \Sigma_\xi = I\\
    A(\beta) = \text{diag}\lrc{\alpha_i(\beta)}V^T\\
    \alpha_i(\beta) = \lrs{\frac{\beta(1-\lambda_i) - 1}{\lambda_i s_i}}^{1/2}\Theta\lrp{\beta - \frac{1}{1-\lambda_i}}
    \end{cases}
\end{equation}
\noindent
Where $\Theta$ is the Heaviside step function and $s_i = [V^T\Sigma_X V]_{ii}$. The matrix $V$ represents a set of eigenvectors with corresponding eigenvalues $\lambda_i$ in the following way:
\begin{equation*}
    \Sigma_X^{-1}\Sigma_{X|Y} V = V \text{diag}\lrc{\lambda_i}\,.
\end{equation*}

The matrix $\Sigma_X^{-1}\Sigma_{X|Y}$ used above also appears in canonical correlation analysis and we therefore refer to it as the ``canonical correlation matrix''. Note that since it is not generally symmetric, the eigenvector matrix $V$ is not generally orthogonal. An important property of the canonical correlation matrix is that its eigenvalues lie within the unit interval; that is, $\lambda_i \in [0,1]$ for all $i$.

The GIB solution \eqref{eq:GIB_sol} is not unique. At a cursory level, this follows from the IB objective function \eqref{eq:IB_objective}, which is a function only of mutual information terms and hence invariant to all invertible transformations on $X$, $\tilde{X}$, and $Y$. However, not all invertible transformations $X \to f(X)$ will leave the joint distributions $P(x,y)$ and $P_\beta(x,\tilde{x})$ Gaussian. It is specifically invertible linear transformations $X \to LX$ (and analogous transformations for $\tilde{X}$ and $Y$) which preserve IB optimality and leave all joint distributions Gaussian. One consequence of this is that $\tilde{X} \to L\tilde{X}$ changes the coarsening parameters $(A,\Sigma_\xi) \to (LA, L \Sigma_\xi L^T) = (A',\Sigma_\xi')$. If $L$ is invertible, then these new parameters also solve GIB. When testing whether a given parameter combination $(A,\Sigma_\xi)$ is GIB-optimal, it is therefore useful to consider the quantity $V^{-1}A^T\Sigma_\xi^{-1}AV^{-T}$, Which is invariant to all invertible linear transformations on $X, \tilde{X}$, and $Y$. 

In this section, we show that solutions to GIB have an exact semigroup structure, wherein two GIB solutions ``chained together'' compose a larger solution which is still optimal. To be more precise, let $P(x,y)$ be jointly Gaussian, then suppose $P_{\beta_1}(\tilde{x}_1|x)$ is IB optimal. Because $P_{\beta_1}(\tilde{x}_1|x)$ is Gaussian under the parameterization $\tilde{X}_1 = A_1 X + \xi_1$, it must also be that $P(\tilde{x}_1,y)$ is jointly Gaussian and thus a valid starting point for a new GIB problem. Taking $\tilde{X}_1$ to be the new input variable, let the second optimal coarsening map be $P_{\beta_2}(\tilde{x}_2|x)$, and parameterize it the same way: $\tilde{X}_2 = A_2\tilde{X}_1 + \xi_2$. Then, we claim, the composition of these two coarsening maps, obtained by integrating the expression $P_{\beta_2}(\tilde{x}_2|\tilde{x}_1)P_{\beta_1}(\tilde{x}_1|x)$ over $\tilde{x}_1$, is also given by a single IB-optimal coarsening $P_\beta(\tilde{x}|x)$ for some $\beta = \beta_2 \circ \beta_1$, where $\circ$ is a binary operator whose explicit form will be provided shortly. We represent this composition schematically with the Markov chain:
\begin{equation}\label{eq:chain}
    Y \leftrightarrow X \xrightarrow{\beta_1} \tilde{X}_1 \xrightarrow{\beta_2} \tilde{X}_2
\end{equation}

To simplify the analysis, we begin by redefining \footnote{To clarify notation: by $X\to L X$, we mean that each instance of $LX$ should be replaced by $X$.} the input variable $X$ by projecting it onto the eigenvectors of the canonical correlation matrix. Assuming that $V$ is full-rank,
\begin{equation*}
    X \to V^TX
\end{equation*}
\noindent
is an invertible linear transformation. Invertibility guarantees that the objective function is unaffected, while linearity guarantees that $P(y,x)$ remains Gaussian. We call this new basis for $X$ the ``natural basis" since after this transformation $\Sigma_X$, $\Sigma_{X|Y}$, and $A_1$ are diagonal. Additionally, after the first compression to $\tilde{X}_1$, the new analogous quantities, e.g. $\Sigma_{\tilde{X}_1}$, $\Sigma_{\tilde{X}_1|Y}$, and $A_2$ will remain diagonal. For the transformation matrices $A_1$ and $A_2$, this fact can be seen by inspecting \eqref{eq:GIB_sol}, while Lemma B.1. in \cite{chechik_information_2005} proves that $\Sigma_X$ and $\Sigma_{X|Y}$ are diagonal. In this new basis, they are given by:
\begin{eqnarray*}
    (\Sigma_X)_{ij} &=& s_i \delta_{ij}\,,\\
    (\Sigma_{X|Y})_{ij} &=& s_i\lambda_i \delta_{ij}\,.
\end{eqnarray*}

We now show that successively applied GIB compression as portrayed in \eqref{eq:chain} composes GIB transformations of greater compression. A more detailed treatment is given in appendix \ref{app:semigroup}. Suppose that $A$ and $\Sigma_\xi$ describe a non-deterministic map $AX + \xi$. From Lemma A.1. in \cite{chechik_information_2005}, this map $(A, \Sigma_\xi)$ is IB-optimal if there exists some $\beta$ such that
\begin{equation*}
    [A^T\Sigma_\xi^{-1} A]_{ij} = \alpha_i^2(\beta)\delta_{ij}\,,
\end{equation*}
\noindent
where $\alpha_i$ is as given in \eqref{eq:GIB_sol}.

Consider two successive maps with bottleneck parameters $\beta_1$ and $\beta_2$, each with unit noise. The composition of these transformations is represented by the pair $(A,\Sigma_\xi)$ $=$ $(A_2A_1, A_2A_2^T + I)$. 
Both $A_1$ and $A_2$ can be computed explicitly using \eqref{eq:GIB_sol}, though $A_2$ is initially given in terms of the statistics $P(\tilde{x}_1,y)$. Using $\tilde{X}_1 = A_1 X + \xi_1$, we thus re-write $A_2$ in terms of the original relevance variable-input variable statistics $P(x,y)$. After this substitution, direct evaluation of $A^T\Sigma_\xi^{-1}A$ shows that $(A_2A_1,A_2A_2^T + I)$ is IB optimal:
\begin{equation*}
    [A_1A_2(A_2^2+I)^{-1}A_2A_1]_{ij} = \alpha_i^2(\beta_2\circ\beta_1)\delta_{ij}\,,
\end{equation*}
\noindent
where $\beta_2 \circ \beta_1$ is the bottleneck parameter of the full, 1-step compression:
\begin{equation}\label{eq:beta_comp}
    \beta_2 \circ \beta_1 = \frac{\beta_2 \beta_1}{\beta_2 + \beta_1 - 1}\,.
\end{equation}

It is important to note that this computation \textit{defines} the binary operator $\circ$. If GIB did not have a semigroup structure, it would not be possible to identify $\circ$ in this manner. Direct computations show that this operator satisfies closure and associativity, and thus furnishes the space in which $\beta$ values live, that is $\mathbb{R}>1$, with a semigroup structure. As bonuses, if we consider $\beta=\infty$ to be an element, we see that it is the identity element. This aligns with the fact that in the limit $\beta\to\infty$, the IB objective \eqref{eq:IB_objective} becomes insensitive to the encoding cost $I(X;\tilde{X})$ and hence no coarsening occurs; $\tilde{X}$ becomes a deterministic function of every component of $X$ which contains information about $Y$. Further, $\circ$ is symmetric. One should be careful to note, however, that the maps $P_{\beta_1\circ\beta_2}(\tilde{x}|x)$ and $P_{\beta_2\circ\beta_1}(\tilde{x}|x)$ need only agree in the overall level of compression achieved, and may otherwise differ since $\tilde{X} \to L\tilde{X}$ is a symmetry.

\subsection{What is the significance of this structure?}

A broad goal of this paper is to explore structural similarities between IB and RG. The semigroup structure present in Wilsonian RG is crucial to its explanation of scaling phenomena, so its presence in GIB is a promising sign. The traditional picture is this: consider the RG transformations $\mathcal{R}_{b_1}$ and $\mathcal{R}_{b_2}$ which rescale length by factors $b_1$ and $b_2$, respectively. Then a fundamental property of $\mathcal{R}$ is that $\mathcal{R}_{b_1}\mathcal{R}_{b_2} = \mathcal{R}_{b_1b_2}$. This structure imposes a strong constraint on the behavior of the flow near a fixed point. If $\sigma$ represents an eigenvector of the Jacobian matrix at the fixed point, then its associated eigenvalue $\lambda_\sigma$ will scale as $b^{y_{\sigma}}$\cite{goldenfeld_lectures_2018}. In short, the semigroup typically allows one to define the critical exponent $y_\sigma$.

The operator $\circ$ we introduced does not immediately lend itself to this sort of analysis. However, we can introduce a function $b(\beta)$ which satisfies $b(\beta_2\circ\beta_1)$ $=$ $b(\beta_2)b(\beta_1)$. By inspection, this function is given by:
\begin{equation*}
    b(\beta) = \frac{\beta}{\beta - 1}\,.
\end{equation*}

This quantity is interesting because it is analogous to the length-rescaling factor found in typical Wilsonian or Kadanoff RG schemes, yet in IB there is no need for a notion of space, and hence rescaling length generally means nothing. Compare this with, for example, a momentum-shell decimation scheme. One identifies the rescaling factor by comparing the new and old UV cutoffs, and so it acquires the meaning of a length-rescaling factor. Here, $b$ is determined entirely by the Lagrange multiplier $\beta$ and the structure of GIB, both of which are defined without deference to an \textit{a priori} existing notion of spatial extent. As discussed in the introduction, connecting IB to RG is attractive, in part, precisely because IB is an information-theoretic framework and does not rely on physical interpretations. Hence this rescaling factor $b$ should be considered an information-theoretic quantity in the same way as $\beta$. 

Can $b(\beta)$ as defined above be used in the same way as the rescaling factor $b$ is used in RG? First, limits of the IB problem involving extremal values of $\beta$ should match intuition about $b$ in an RG context. Indeed, for $\beta\to\infty$, the zero-coarsening limit, $b(\beta) \to 1$. Next, by the data processing inequality, at $\beta = 1$ the optimal IB solution is degenerate with complete coarsening, i.e. $\tilde{X}$ becomes independent of $X$. Correspondingly, the limit $\beta\to1$ gives $b(\beta)\to \infty$. Next, let us recall the scope of the GIB framework. GIB makes statements only about completely Gaussian statistics, so no anomalous scaling will appear, and thus a discussion of critical exponents is hard to motivate. Second, GIB is defined for finite-dimensional $X$ and $Y$, so we cannot simply connect it to, say, momentum-shell Wilsonian RG, which only makes statements about infinite systems. Finally, and related to the last point, we have not identified yet what the analogous ``model space'' is in the context of IB, or how an optimal GIB map could represent an RG transformation in that space. This will be the subject of the next section, where we show that the non-deterministic nature of IB coarsening aligns exactly with existing soft-cutoff RG methods.

Whether or not this analysis helps to formally connect IB and RG, it is interesting to ask whether other IB problems exhibit semigroup structure. One could imagine, for example, that a series of high-$\beta$ compression steps (low-compression limit) might be easier than one large compression step. If this is the case, IB problems with semigroup structure may benefit from an iterative chaining scheme similar to the one we present here. One possible application of this structure is the construction of feed-forward neural networks with IB objectives. If the IB problem in question has semigroup structure, then the task of training the entire network can be reduced to training the layers one-by-one on smaller (higher-compression) IB problems. This has benefits in biological systems, such as biochemical and neural networks, where processing is often hierarchical, likely as a result of underlying evolutionary and developmental constraints. Biological systems are also shaped by their output behavior, which sets a natural relevance variable in the arc from sensation to action. 

\section{Structural similarities between IB and NPRG}

\subsection{\label{sec:soft_is_nondeterminsitic}Soft-cutoff NPRG is a theory of non-deterministic coarsening}

The renormalization group is not a single coherent framework, but rather a collection of theories, computational tools, and loosely-defined motifs. As such, it is probably not possible to succinctly define RG on the whole. A common theme, at least, is that RG techniques describe how the effective model of a given system changes as degrees of freedom are added or removed. The modern view of RG theory, which is largely due to Wilson \cite{wilson_renormalization_1971, wilson_renormalization_1971-1,wilson_renormalization_1974,wilson_critical_1972,fisher_renormalization_1998} and Kadanoff \cite{kadanoff_scaling_1966}, concerns itself with the removal of degrees of freedom through a process known as decimation, in which a thermodynamic quantity (typically the partition function) is re-written by performing a configurational sum or integral over a subset of the original modes. Here, even before discussing rescaling and renormalization, we must make procedural choices. To begin, one must specify the subset of degrees of freedom which are to be coarsened off. In theories where modes are labelled by wavenumber or momentum, one typically establishes a cutoff and decimates all modes with momentum above it. As a result, those modes are completely removed from the system description, and their statistics are incorporated into the couplings which parameterize the new effective theory. Another consideration is the practicality of carrying out such a procedure. If the model in consideration can be expanded in a perturbation series about a Gaussian model, and if the non-Gaussian operators are irrelevant or marginal under the flow, then this analysis is amenable to perturbative RG. However, this is often not the case, for example in systems far from their critical dimension, or in non-equilibrium phase transitions, where there may not even be critical dimensions \cite{canet_quantitative_2004, canet_nonperturbative_2005}.

%

In non-perturbative RG (NPRG) approaches, the need for a perturbative treatment is removed by working from a formally exact flow equation at the outset. The first such treatment was put forth in 1973 by Wegner and Houghton, who used Wilson's idea of an infinitesimal momentum-shell integration to derive an exact flow equation for the full coarse-grained Hamiltonian \cite{wegner_renormalization_1973}. Because this equation describes the evolution of the Hamiltonian for every field configuration, this and other NPRG flow equations are called integro-differential equations, and the NPRG is sometimes referred to as the functional renormalization group (FRG). Later, Wilson and Kogut \cite{wilson_renormalization_1974}, as well as Polchinski \cite{polchinski_renormalization_1984}, proposed new NPRG flow equations in which the cutoff was not described explicitly through a literal demarcation between included and excluded modes, but instead through non-deterministic coarsening, so that the effective Hamiltonian satisfies a functional generalization of a diffusion equation \footnote{This interpretation is explicitly presented in the Wilson-Kogut paper. Given that their approach is formally equivalent with the one put forth by Polchinski, the interpretation should apply to that framework as well.}. These approaches were introduced, at least in part, as a response to difficulties \footnote{Under the sharp cutoff construction, some issues include the generation of non-local position-space interactions \cite{wilson_renormalization_1974}, unphysical nonanalyticities in correlation functions, and the need to evaluate ambiguous expressions such as $\delta(x)f(\Theta(x))$ where the function $f$ is not known \textit{a priori} \cite{kopietz_introduction_2010}.} that arise from the sharp cutoff in the Wegner-Houghton construction. Correspondingly, the Wilson-Polchinski FRG approach can be thought to give a soft cutoff, where modes can be ``partially coarsened".

The most common NPRG approach in use today was first described in 1993 by C. Wetterich \cite{wetterich_exact_1993}. Like the Wilson-Polchinski NPRG, the Wetterich approach uses a soft cutoff, but the objects computed by this framework are fundamentally different. Instead of computing the effective Hamiltonian of modes which are below the cutoff, the Wetterich framework computes the effective free energy of modes \textit{above} the cutoff. For this reason, we say that the Wilson-Polchinski framework is UV-regulated and Wetterich is IR-regulated. Yet, despite this difference in perspective, the Wetterich formalism still describes the flow effective models make from their microscopic to macroscopic pictures. In this section, we will explore how the soft-cutoff construction is related to a notion of non-deterministic coarsening, and in turn, the information bottleneck framework. An in-depth discussion of the philosophy and implementation of NPRG techniques would be distracting, so we instead refer the reader to a number of good references on the topic \cite{bagnuls_exact_2001, berges_non-perturbative_2002, delamotte_introduction_2012, kopietz_introduction_2010}. 

So far we have not explained how one actually imposes a soft-cutoff scheme. We begin by examining the Wetterich setup, in which one writes the effective (Helmholtz) free energy at cutoff $k$:
\begin{eqnarray*}
    W_k[J] &=& \log \int \DD \chi \exp \bigg[- S[\chi] - \Delta S_k[\chi] \\ 
    & & + \sum_a\int \dd^d x\, J_a(x) \chi_a(x) \bigg]\,.
\end{eqnarray*}

The bare action, given by $S$, is the microscopic theory which is known \textit{a priori}. The source $J$ allows us to take (functional) derivatives of this object to obtain cumulants (connected Green's functions). The remaining term $\Delta S_k[\chi]$ is known as the deformation, and it is this term which enforces the cutoff. It is written as a bilinear in $\chi$:
\begin{eqnarray}
    \Delta S_k[\chi] &=& \frac{1}{2}\sum_{ab}\int \dd^d x\, \dd^d y\, [R_{k}]_{ab}(x,y) \chi_a(x)\chi_b(y)\,.
\end{eqnarray}

For compactness, we will often resort to a condensed notation and express integrals instead as contraction over suppressed continuous indices. For example, the deformation may be re-written:
\begin{equation*}
    \Delta S_k[\chi] = \frac{1}{2}\, \chi^\dagger R_k \chi \,.
\end{equation*}

The kernel (matrix) $R$ is known as the regulator, and it controls the ``shape'' of the cutoff. Almost always, it is chosen to be diagonal in Fourier basis so that the cutoff $k$ has the interpretation of a wavenumber or momentum. The resulting Fourier-transformed regulator $R_k(q)$ has some freedom in its definition, but it must satisfy the following properties \cite{litim_optimized_2001}:
\begin{enumerate}
    \item $\lim_{q^2/k^2 \to 0} R_k(q) > 0$ 
    \item $\lim_{k^2/q^2 \to 0} R_k(q) = 0$
    \item $R_k(q) \to \infty \quad  \forall q\quad\text{ as }\quad k \to \infty$
\end{enumerate}

These constraints guarantee that the deformation acts as an IR cutoff. The first condition increases the effective mass of low-momentum modes and suppresses their contribution to the effective free energy. The second ensures that modes with high momentum ($q^2 > k^2$) are left relatively unaffected, and contribute more fully to $W_k$. The third condition ensures that the so-called ``effective action,'' defined as 
\begin{equation*}
    \Gamma_k[\varphi] = J^\dagger \varphi - W_k[J] - \Delta S_k[\varphi]\,,
\end{equation*}
\noindent
approaches the bare action (or Hamiltonian, as the case may be) in the limit $k\to\infty$. Here, the order parameter $\varphi$ is given by $\delta W_k[J]/\delta J^\dagger $. Because of this construction, the second regulator property also ensures that in the limit $k\to0$, the deformation $\Delta S_k$ disappears, and the effective action $\Gamma_k$ becomes the Legendre transform of $W[J]$. This functional $\Gamma_{k=0}$ is known in many-body theory as the 1PI generating functional, and in statistical mechanics as the Gibbs free energy. In the Wetterich formalism, one is generally interested in computing the flow of $\Gamma_k$ because of these useful boundary conditions. 

To see how this approach is related to non-deterministic coarsening, we will connect it to a soft-cutoff UV-regulated approach, also put forth by Wetterich, which is formally equivalent to the Wilson-Polchinski framework. We begin with the following expression defining the average action $\Gamma^{\text{av}}_k[\tilde{\chi}]$, taken directly from the paper \cite{wetterich_average_1993}, with only a slight change in notation:
\begin{equation*}
    -\Gamma_k^{\text{av}}[\tilde{\chi}] = \log \int \DD \chi\, P_k[\tilde{\chi}|\chi] \exp(-S[\chi])\,,
\end{equation*}
\noindent
where we refer to this functional $P_k[\tilde{\chi}|\chi]$ as the coarsening map. If we were interested in performing deterministic coarsening, i.e. one involving a hard cutoff, the coarsening map would be something like a delta-function $\delta(\tilde{\chi} - \Phi_k[\chi])$ for some functional $\Phi_k$. However, in all soft-cutoff UV-regulated approaches, this distribution is Gaussian in $\tilde{\chi}$
\begin{equation}\label{eq:NPRG_coarsening}
    P_k[\tilde{\chi}|\chi] = \exp\lrs{-\frac{1}{2} \lrp{\tilde{\chi}- A_k\chi}^\dagger \Delta_k^{-1}\lrp{\tilde{\chi}-A_k\chi} - C_k} \,.
\end{equation}

In principle, given the coarsening parameters $A_k$ and $\Delta_k$ for all $k$, the exact flow equation for $\Gamma^{\text{av}}_k$ is determined. Wetterich gives explicit choices for these parameters, while Wilson and Polchinski independently give their own (though in slightly different fashion). The term $C_k$ is a normalizing constant which is essentially unimportant to the remainder of our discussion.

Now we connect the IR and UV approaches to show that they are complementary, and in some sense, equivalent. In particular, suppose we know $P_k[\tilde{\chi}|\chi]$ for all $k$. Then, from this single object, one can construct both the IR-regulated and UV-regulated flows. This should make intuitive sense; the IR-regulated part tracks the thermodynamics of the already-integrated modes, while the UV-regulated part tracks the model of the unintegrated modes. This can all be seen clearly by writing out the full sourced partition function $Z[J]$ and invoking the normalization of the coarsening map.
\begin{eqnarray}
    Z[J] &=& \int \DD \chi\, \exp\lrp{-S[\chi] + J^\dagger \chi} \nonumber\\ 
    &=& \int \DD \tilde{\chi}\, \DD \chi\, P_k[\tilde{\chi}|\chi]\exp{\lrp{-S[\chi]+J^\dagger\chi}} \nonumber\\ 
    &\sim& \int \DD \tilde{\chi}\, \exp \lrp{-\frac{1}{2}\tilde{\chi}^\dagger \Delta_k^{-1}\tilde{\chi} + W_k[J + \tilde{J}[\tilde{\chi}]]}  
    \label{eq:Z_sourced}
\end{eqnarray}

In the final expression, the normalizing constant $C_k$ has been dropped. Readers familiar with the Polchinski formulation will immediately recognize $W_k[\tilde{J}[\tilde{\chi}]]$ as the effective interaction potential. However, the argument to this potential is shifted by the source $J$, which therefore enters nonlinearly, unlike in Polchinski's approach. This difference is due to the fact that we define a flow for each initial source configuration, instead of adding a linear source term to the vacuum flow.

To arrive at \eqref{eq:Z_sourced} above, we had to define the effective field-dependent source $\tilde{J}$ and identify a suitable deformation term in $P_k[\tilde{\chi}|\chi]$. By directly substituting \eqref{eq:NPRG_coarsening}, one can see that
\begin{equation*}
    \tilde{J}[\tilde{\chi}] = A_k^\dagger \Delta_k^{-1} \tilde{\chi}\,.
\end{equation*}
\noindent
and
\begin{equation} \label{eq:IB_deformation}
    \Delta S_k[\chi] = \frac{1}{2}\chi^\dagger A_k^\dagger \Delta_k^{-1} A_k \chi\,,
\end{equation}

As promised, the existence of a family of distributions $P_k[\tilde{\chi}|\chi]$ with a known parameterization $(A_k, \Delta_k)$ allows us to define an IR regulator scheme, and therefore compute the NPRG flow both above and below the cutoff. The deformation term $\Delta S_k$ ultimately came from the $\chi^2$ term present in the coarsening map, which could be interpreted as a free energy. We also identify immediately that the IR regulator $R_k$ corresponding to a given choice of coarsening map is given by $A_k^\dagger \Delta_k^{-1}A_k$. 

We will next use this viewpoint to introduce information bottleneck into the discussion. In particular, we will associate the coarsening map $P_k[\tilde{\chi}|\chi]$ with the IB coarsening map $P_\beta(\tilde{x}|x)$ and examine some consequences. This discussion comes with some restrictions. Firstly, one should note that all soft-cutoff NPRG frameworks, regardless of the structure of the microscopic action, assume a Gaussian coarsening map. With a non-Gaussian $P_k[\tilde{\chi}|\chi]$, the flow may still be defined, but it will not, in general, satisfy any known exact flow equations. This is easiest to see in the IR Wetterich formalism, since a non-Gaussian $P_k$ would yield a $\Delta S_k[\chi]$ which is no longer bilinear in $\chi$, and hence one could not write the flow equation in terms of the exact effective propagator, as it usually is. Indeed, the more general $\Delta S_k[\chi]$ could have terms at arbitrarily high order in $\chi$, and thus require arbitrarily high-order derivatives of $\Gamma_k$ in the flow equation. So, while it is not impossible to seriously consider non-Gaussian $P_k[\tilde{\chi}|\chi]$, it is certainly inadvisable without good reason. 

With this in mind, we must also note that IB has an exact solution involving Gaussian $P_\beta(\tilde{x}|x)$, but only when the variables $X$ and $Y$ are jointly Gaussian. By analogy, this restricts us to discussing theories where the bare action $S[\chi]$, or perhaps more accurately, the bare Hamiltonian $\mathcal{H}[\chi]$ contains only linear and bilinear terms in $\chi$. While everything presented above holds for general $S$, everything that follows will be totally Gaussian so that IB optimality can be exactly satisfied. Finally, note that IB may not be well-defined for infinite-dimensional random variables such as fields, so our scope is further limited to finite-dimensional multivariate Gaussian distributions of classical variables.

\subsection{\label{sec:GIB_regulator}The Gaussian IB regulator scheme}

In the last section, we briefly introduced soft-cutoff NPRG approaches and argued that both UV- and IR-regulated flows can be defined given a family of Gaussian coarsening maps $P_k[\tilde{\chi}|\chi]$. Broadly, we aim to show in this paper that IB and RG can be connected by identifying this map with the IB-optimal coarsening map $P_\beta(\tilde{x}|x)$. By this we do not mean to say that the family of maps produced by IB are the ``correct'' starting point for NPRG. Instead, we simply note that IB-optimality is a constraint one could impose on the coarse-graining scheme. Assuming we do so, what characteristics does the IB-RG scheme carry? Using the exact solution to GIB and Eq. \eqref{eq:IB_deformation}, we identify the regulator, or soft-cutoff scheme, required by IB optimality for some known initial statistics $P(x,y)$
\begin{equation}\label{eq:R_IB_def}
    \lrs{R_\beta^{(\text{\IB})}}_{ij} = \frac{\beta - \beta_i}{s_i (\beta_i - 1)} \Theta(\beta - \beta_i)\delta_{ij}
\end{equation}

Here the $\beta_i$ are critical bottleneck values, indexed by the components of the so-called ``natural'' basis, which is found by diagonalizing the canonical correlation matrix $\Sigma_X^{-1}\Sigma_{X|Y}$ as discussed in section \ref{sec:semigroup}. The critical bottleneck values $\beta_i$ are given by $(1-\lambda_i)^{-1}$. If $V$ is the matrix of right eigenvectors of this matrix, then $s_i$ is given by $[V^T\Sigma_X V]_{ii}$. Notice also that this regulator is diagonal in natural basis. $\Theta$ denotes the Heaviside step function.

In the typical context, $R$ is diagonalized by a Fourier transform, and thus it represents a cutoff in wavevector or momentum. Here this notion is generalized, and instead of identifying a cutoff wavenumber $k$, we should consider the cutoff to be of information-theoretic origin, and fundamentally defined by $\beta$. Consequently, the degree to which the mode labelled by $i$ is coarsened should be found by comparing its corresponding critical value $\beta_i$ to the cutoff $\beta$. As such, we can essentially make the replacements $k^2 \to \beta$ and $q^2 \to \beta_i$, with the caveat that $\beta$ and $\beta_i$ should approach unity as $k^2$ and $q^2$ go to zero.

\begin{figure}[b]
\hspace{-5mm}
\includegraphics[width = 0.48\textwidth]{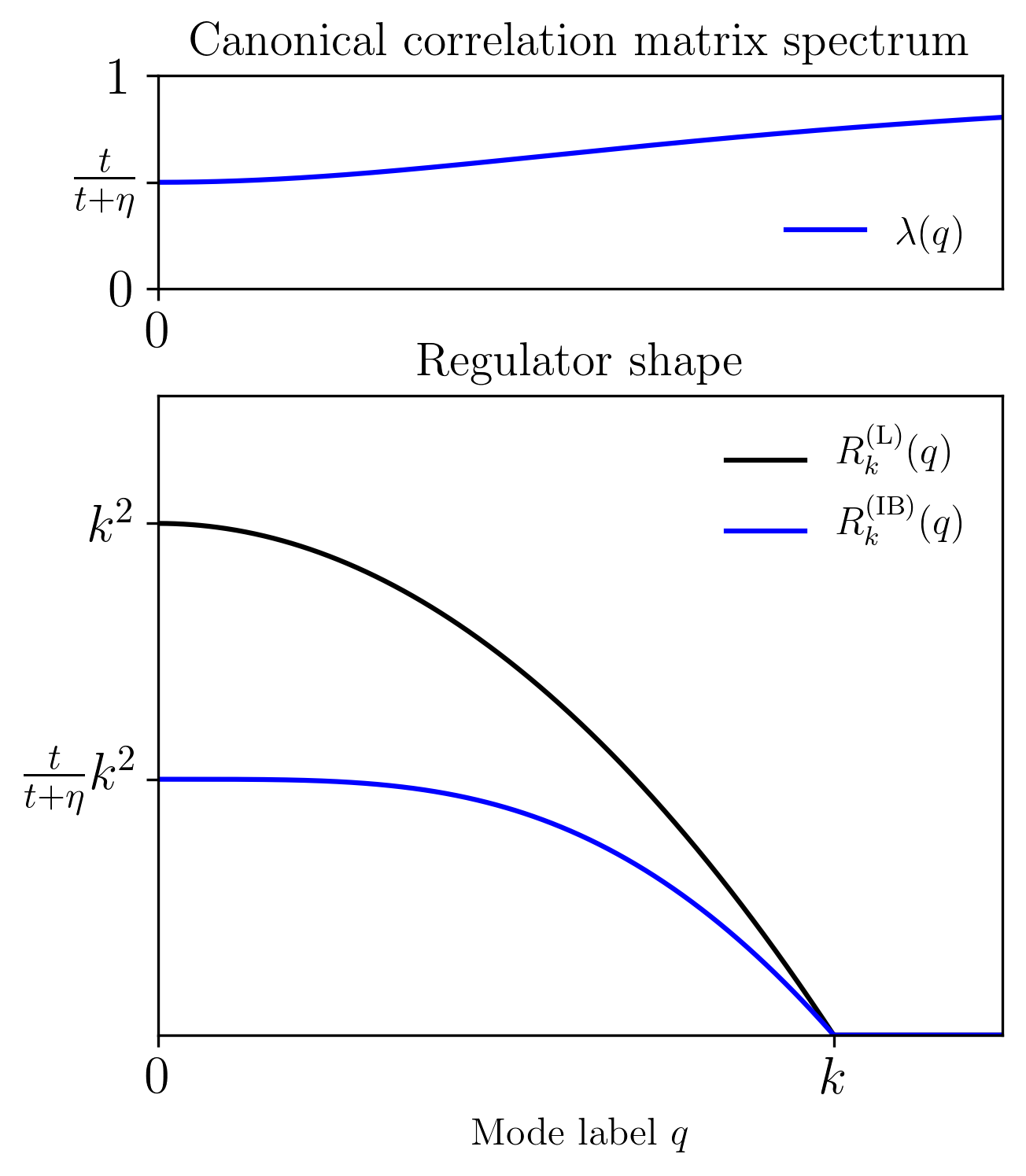}
\caption{\label{fig:IB_regulator_shape}IR Regulators compared between Litim and IB schemes. The IB problem depicted here is from the toy model discussed in section \ref{sec:not_always_fourier} for the simple case where the collective modes selected by IB are Fourier and the disorder correlation has no dispersion ($\eta$ is constant). \textit{Top}: Eigenvalues of the canonical correlation matrix $\Sigma_X^{-1}\Sigma_{X|Y}$ as a function of label $q$, which may be interpreted as a wavevector magnitude. Modes with smaller eigenvalue can be thought to carry more information about $Y$. \textit{Bottom}: Regulator values as a function of cutoff $k$ and mode label $q$ for the Litim scheme \eqref{eq:R_Litim_def} and the IB scheme (black and blue, respectively). }
\end{figure}

In Figure \ref{fig:IB_regulator_shape}, we plot $R^{(\IB)}$ obtained from the first toy model presented in section \ref{sec:not_always_fourier} and compare it against the well-known Litim regulator \cite{litim_optimized_2001}, denoted $R^{(\text{L})}$ and given in Eq. \eqref{eq:R_Litim_def}. Ignoring for now the particulars of the model, we point out that the IB and Litim regulators appear qualitatively similar, and for fixed parameters $t$ and $\eta$, all limits involving $q$ and $k$ satisfy the regulator scheme requirements. Moreover, we see that the NPRG and IB notions of mode relevance are in agreement. Smaller canonical correlation eigenvalues $\lambda$ (top plot) correspond to collective modes which get integrated out later in the flow. This is reflected in the structure of the soft cutoff, which increasingly suppresses fluctuations as $q\to0$.

Is it okay to take \eqref{eq:R_IB_def} seriously as an IR regulator scheme? Let us attempt to compare with the conditions outlined in the last section. The typical interpretation of the first requirement on $R$ is that the lowest energy modes should be given extra mass by the regulator so that they are ``frozen out'' of the configurational integral. In fewer words, there should not be soft modes in intermediate stages of the flow. By analogy, it must be true that $(R_\beta)_{11} > 0$, where we take $\beta_1 = \min_i \beta_i$ to represent the most ``relevant'' mode (in the IB sense). Indeed, for all $\beta$, this is satisfied by \eqref{eq:R_IB_def}. Next, $R$ must vanish for the $i^{\text{th}}$ mode when the cutoff $\beta$ is taken sufficiently far below $\beta_i$. Because of the step function, this is satisfied. Finally, each diagonal component $(R_\beta)_{ii}$ should diverge as $\beta \to \infty$ so that at zero compression, only the saddle point configuration of the microscopic theory contributes to the generating function, or whichever thermodynamic potential we are interested in. If $\beta_i$ are all finite, then this limit holds as well \footnote{Two edge cases may appear important; $\beta_i\to 1$ and $\beta_i\to \infty$. Because $\beta_i = (1-\lambda_i)^{-1}$, these correspond to limits where a mode $[V^TX]_i$ is deterministically related to $Y$ or completely independent of $Y$, respectively. Complete deterministic dependence between $X$ and $Y$ should not be considered without modification, and in the case of independence, those modes may be removed as a formality.}. 

Because it satisfies all of the properties required of a typical regulator in a soft-cutoff scheme, we call \eqref{eq:R_IB_def} the ``IB regulator" and denote it $R^{(\IB)}$. This identification has some interesting consequences, which will be explored in the coming section. One particularly striking feature is that the cutoff scheme is now parameterized by the family of distributions $P(x,y)$. In IB theory, these distributions formalize the notion of ``important features'' of $X$ implicitly through its correlations with $Y$. This means that the RG scheme selected by a given set of IB solutions will \textit{not} favor, for instance, ``long distance modes'' unless $P(x,y)$ is chosen to enforce that. Instead, the analogue of long distance modes are those modes which have the most information about $Y$. In section \ref{sec:not_always_fourier} we will attempt to clarify this by calculating the IB regulator explicitly in a simple, familiar context.

\section{Consequences and interpretations of the correspondence}

\subsection{\label{sec:BA}The Blahut-Arimoto update scheme may displace the flow-equation description}

The apparent goal of Information Bottleneck is to identify the coarsening map $P_\beta(\tilde{x}|x)$ for some set of $\beta$ values. This seems to align poorly with the problem statement and goals of NPRG, in which the coarsening map $P_k[\tilde{\chi}|\chi]$ is taken as the starting point and used to derive the flow equations. Is it really true that solving IB only gets us to the starting point of an RG scheme, after which we still need to ``do the RG part?'' In this section, we investigate one way to resolve this dissonance by noting that the quantities one would usually consider to be the results of the NPRG flow can be used to parameterize $P_\beta(\tilde{x}|x)$ itself. From this viewpoint, one may organize the computation around a set of self-consistent update equations instead of a set of flow equations.

The general IB problem can be solved, in principle, by iterating what is known as the Blahut-Arimoto procedure, which is borrowed from rate distortion theory in a more general context \cite{tishby_information_2000}. This procedure relies on the fact that when $P_\beta(\tilde{x}|x)$ is IB optimal, it satisfies the following condition:
\begin{equation*}
    P_\beta(\tilde{x}|x) = Z_\beta(x)^{-1} P_\beta(\tilde{x})\exp\lrp{-\beta D_{\text{KL}}[P(y|x)||P_\beta(y|\tilde{x})]} \,,
\end{equation*}
\noindent
Where everything on the RHS is to be considered a function of $P_\beta(\tilde{x}|x)$ through
\begin{eqnarray*}
    P_\beta(\tilde{x}) &=& \int \dd x\, P_\beta(\tilde{x}|x)P(x)\,, \\ 
    P_\beta(y|\tilde{x}) &=& \frac{1}{P_\beta(\tilde{x})}\int \dd x\, P(y|x) P_\beta(\tilde{x}|x)P(x) \,.
\end{eqnarray*}

The function $Z_\beta(x)$ normalizes $P_\beta(\tilde{x}|x)$ and therefore also depends on $P_\beta(\tilde{x}|x)$ through the above equations.

In brief, the BA procedure entails taking an estimate for $P_\beta(\tilde{x}|x)$, plugging it into the IB optimality criterion above, then iterating until satisfactory convergence. In this way, we say that $P_\beta(\tilde{x}|x)$ is self-determined. This procedure is practically very difficult---if not impossible---for distributions of multivariate continuous variables in general. However, in the case of GIB, we can parameterize the distributions then use Gaussian integral identities to update these parameters exactly. Chechik et al. \cite{chechik_information_2005} carry out this procedure in terms of the matrices $A$ and $\Sigma_\xi$, used to define $\tilde{X} = AX + \xi$. We repeat this computation but instead parameterize the update equation using $\Sigma_{\tilde{X}}$, $\Sigma_{X|\tilde{X}}$, and $\Sigma_{X\tilde{X}}$. The first two of these represent objects of interest in the UV- and IR-regulated parts of the NPRG scheme, respectively. The third quantity, $\Sigma_{X\tilde{X}}$ carries information about how the IR degrees of freedom $\tilde{X}$ are coupled to the original, UV variables $X$. In a very condensed form, the BA update equations in this parameterization read:
\begin{eqnarray} \label{eq:BA_update}
    \Sigma'_{X|\tilde{X}} \,&=&\, [\Sigma_X^{-1} + \beta^2 B^T \Sigma'_{\tilde{X}|X}B]^{-1}\,,\\
    \Sigma'_{\tilde{X}} \,&=&\, [\Sigma'^{-1}_{\tilde{X}|X} - \beta^2 B \Sigma_{X|\tilde{X}} B^T]^{-1}\,\nonumber, \\
    \Sigma'_{X\tilde{X}} &=& \beta \Sigma'_{X|\tilde{X}}B^T\Sigma'_{\tilde{X}}\,\nonumber.
\end{eqnarray}
\noindent
where both $B$ and $\Sigma'_{\tilde{X}|X}$ can be expressed in terms of $\beta$, $P(x,y)$ and the current estimate for the parameterization of $P(t,x)$. The full expressions are complicated and given fully in appendix \ref{app:BA_updates}. Note that $\Sigma_{X|\tilde{X}}$ represents the IR-regulated flow; it is directly analogous to the effective propagator $G_k$ in the Wetterich formalism. In other words, given that we are only looking at Gaussian statistics, the function $W_\beta(J)$ (or $\Gamma_k$) can be simply reconstructed from $\Sigma_{X|\tilde{X}}$. Next, $\Sigma_{\tilde{X}}$ represents the UV-regulated part, since the probability distribution describing $\tilde{X}$ can be reconstructed from it. 

We reiterate that this self-consistent updating scheme comes from IB optimality, written in terms of objects we would usually calculate in NPRG. The idea of a self-consistent updating scheme which determines the IR-regulated statistics and UV-regulated dynamics simultaneously is interesting. In addition to essentially replacing the flow-equation description, it is very non-perturbative in nature. However, it seems wrong that imposing a constraint on $P(\tilde{x}|x)$ should make anything easier, especially given the fact that IB enforces a goal which is only sometimes aligned with the typical goals of RG analysis. A natural question, then, is whether IB has actually provided any new leverage. More precisely, if we really have given up the flow equation in favor of a self-consistency scheme, does this new scheme actually help to calculate the objects of interest as the flow equation usually would? If so, why would IB optimality be necessary?

In the case of general, i.e. non-Gaussian $P(x,y)$, the integration 
\begin{equation*}
    \int \dd x\, P(y|x) P(x|\tilde{x})
\end{equation*}
\noindent
can't be carried out directly. This is equivalent to the statement that at (and below) intermediate values of $k$ in NPRG, $W_k[J]$ can't be directly computed from its integral representation. The whole point of Wilsonian RG is to get around this integration step by connecting $W_k$ to $W_{k\to\infty} = 0$ by invoking a known flow equation. So, to answer our question, the IB update scheme may actually provide the same leverage, but only if \textit{(1.)} we can represent the BA procedure parametrically, and \textit{(2.)} the derivation of that parametric representation does not require the explicit marginalization of $x$ to obtain $P(y|\tilde{x})$. The updates we present above for the fully Gaussian problem satisfy the first requirement, but fail the second since we explicitly carried out Gaussian integrals over $x$ in the derivation. It is therefore unclear at this point whether some structure in IB could allow us to estimate $P(y|\tilde{x})$ parametrically, which seems to be a prerequisite for the utility of a more general IB-RG framework in which IB is exactly enforced. Finally, we note that these conditions are necessary, but not sufficient, since further integration steps may be required to complete the BA update, for example in computing $D_{KL}[P(y|x)||P(y|\tilde{x})]$ and going from an updated $P(x,\tilde{x})$ back to the moments of $P(x|\tilde{x})$.

In principle, the self-consistent structure imposed by IB-optimality obviates the need for a traditional cutoff/flow equation description. However, the opposite is also true: if the cutoff scheme and flow equation are known, then the self-consistency conditions are displaced. Because GIB is exactly solvable, we are able to examine both approaches here. In Eq. \eqref{eq:R_IB_def}, we present a soft cutoff scheme which arises from the constraint of GIB-optimality, but it is given in terms of quantities which have no physical context, and so it is hard to say \textit{a priori} how it relates to existing cutoff schemes structurally. In the next section, we consider a toy model which provides this physical context and therefore affords us a glimpse into how IB-optimal NPRG schemes differ structurally from those already employed.

\subsection{Collective modes are not always Fourier: a minimal example} \label{sec:not_always_fourier}
In the Wetterich NPRG, the cutoff is enforced through a deformation $\Delta S_k[\chi] = \frac{1}{2}\chi^\dagger R_k \chi$ added to the bare action or Hamiltonian. In section \ref{sec:GIB_regulator}, we identified this structure as the free energy of a Gaussian coarsening map from the bare degrees of freedom $\chi$ to some compressed representation $\tilde{\chi}$. We then defined the IB regulator through the deformation produced by the map solves the Gaussian Information Bottleneck problem, and showed that it satisfies the various ``design'' constraints traditionally placed upon it. An immediate consequence of this construction is that the regulator design space is now parameterized by the joint distributions $P(x,y)$ which define the starting point of IB, and for many such distributions, the preferred basis selected by IB will look nothing like Fourier modes. Of course, for finite systems not organized in a lattice, this is unsurprising; the Fourier basis will not exist in any familiar sense. However, for practitioners of NPRG, it may cause discomfort to consider a regulator $R_{v(\beta)}(u)$ in which the numbers $v$ and $u$ do not represent radii in momentum space. In contrast, for the majority of applications, the standard cutoff scheme is provided by the Litim regulator
\begin{equation}\label{eq:R_Litim_def}
    R_k^{(\text{L})}(q,q') = \delta^d(q-q') (k^2 - q^2) \Theta(k^2 - q^2)\,,
\end{equation}

Which should be interpreted as a soft momentum-space cutoff. The Litim regulator sees widespread use both because it is optimized to give good convergence properties in certain contexts \footnote{The Litim regulator was introduced in the context of NPRG analysis of the $O(N)$ model. Its favorable or ``optimal'' characteristics are manifested through improved convergence properties of so-called ``threshold functions,'' which constitute a frequently encountered class of momentum integrals involving the cutoff.}, and because its simple form often leads to analytically expressible flow equations (after appropriate truncation procedures) \cite{litim_optimisation_2000, litim_optimized_2001}.

The IB regulator $R^{(\text{IB})}_\beta$ given in \eqref{eq:R_IB_def} does not manifestly have any such nice qualities, and in the general case may be difficult to interpret. In this section, we calculate $R^{(\text{IB})}_\beta$ explicitly in a trivial statistical field theory problem to explore its structure in a familiar context and address some of its non-intuitive features. For our model, we consider a real scalar field $\chi(x)$ in $d$ dimensions at equilibrium and finite temperature $k_BT=1$. This fluctuating field will serve as the ``input variable'' $X$. We also add a disordered source field $h(x)$ which will serve as the ``relevance variable'' $Y$. 
\begin{equation} \label{eq:hamiltonian_conditional}
    \mathcal{H}[\chi|h] = \int \dd ^dx \lrc{\frac{1}{2} \chi(x) (t - \nabla^2) \chi(x) - h(x)\chi(x)}
\end{equation}
\noindent
We also give Gaussian statistics to the disorder:
\begin{eqnarray}
    \overline{\mathcal{A}[h]} &=& \det(2\pi H)^{-1/2} \int \DD h\,\mathcal{A}[h] \times \label{eq:disorder_avg}\\ 
    & &\exp \lrp{-\frac{1}{2} \int \dd^d x_1 \dd^d x_2\, h(x_1) [H^{-1}](x_1,x_2)h(x_2)}\nonumber
\end{eqnarray}
\noindent
In our condensed notation, the above equations are re-expressed:
\begin{eqnarray*}
    \mathcal{H}[\chi|h] &=& \frac{1}{2}\,\chi^T G_0^{-1} \chi - h^T\chi\\
    \log P[h] &\sim& -\frac{1}{2}\,h^TH^{-1}h
\end{eqnarray*}

Together, the Boltzmann weight $\mathcal{H}[\chi|h]$ and the distribution $P[h]$ describing the disorder statistics constitute a joint distribution $P[\chi,h]$ which is jointly Gaussian and thus---momentarily casting aside worries about the continuously infinite-dimensional random variables---a valid starting point for GIB. From the IB standpoint, the goal would usually be to construct a coarsened field $\tilde{\chi}(x)$ which discards some information about $\chi$ while encoding as much as possible about the statistics of $h$. However, the goal here is not to discuss $\tilde{\chi},$ but rather to better understand the NPRG cutoff scheme that IB imposes as a consequence of this starting point. Since we have assumed a canonical form for the bare Green's function $G_0^{-1}$ and the source term is $h\cdot \chi$, the only remaining control over $P[\chi,h]$ is the two-point correlation of $h$:
\begin{equation*}
    \overline{h(x_1)h(x_2)} = H(x_1,x_2)
\end{equation*}

To explore different forms of $R_\beta^{(\text{IB})},$ we therefore consider three different constructions of $H$. First, we choose $h$ to be totally uncorrelated at different points, with a constant variance at each point. Second, we choose $H$ diagonal in Fourier basis, but with some dispersion that adds position-space correlations. In both of these first examples, we will arrive at regulators with momentum-space cutoffs. It is the goal of the third case to present an $H$ which is not diagonal in momentum basis, thereby introducing a non-momentum cutoff structure. 

\begin{figure*}
\includegraphics[width = 0.9\textwidth]{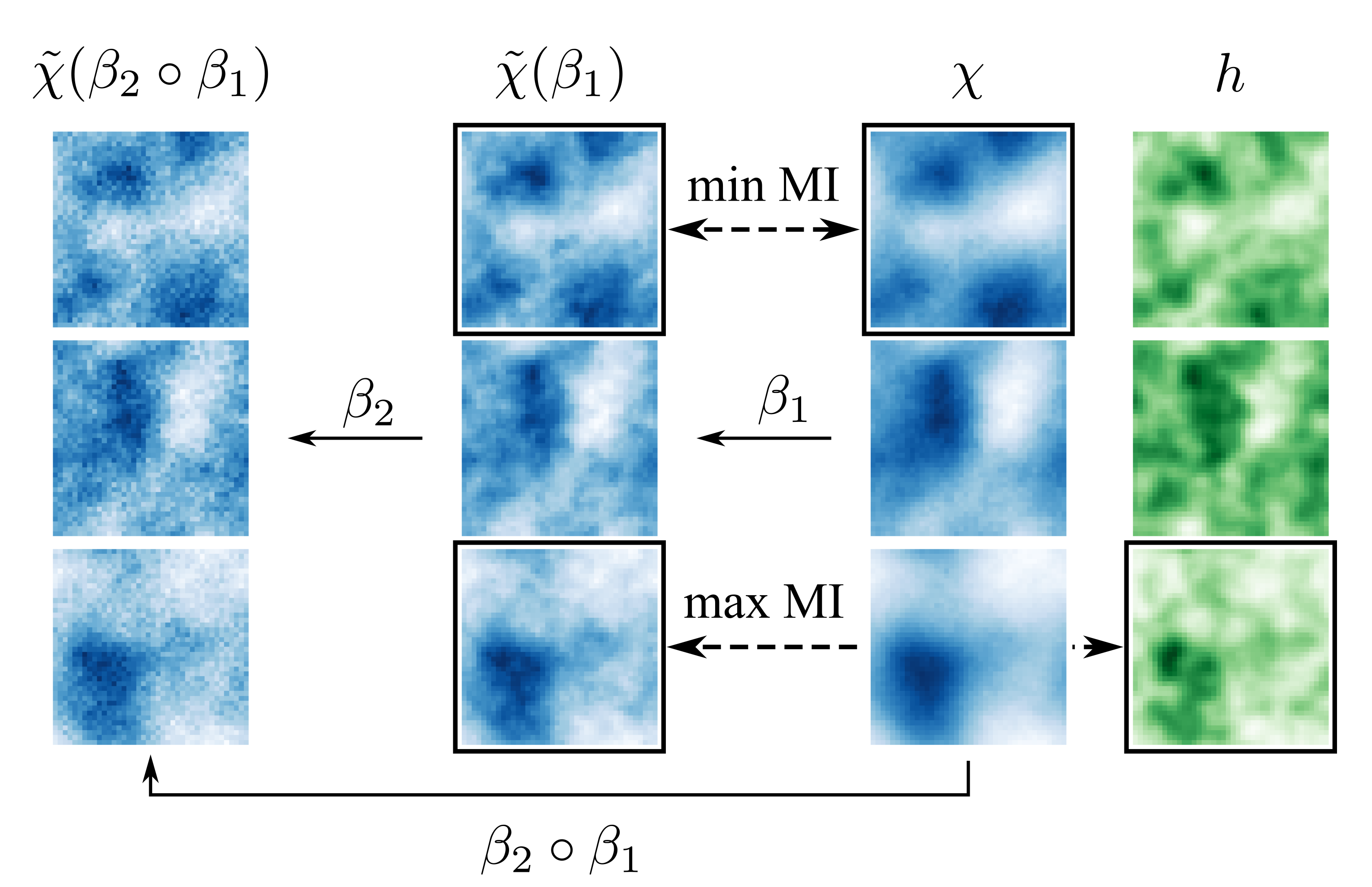}
\caption{\label{fig:toy_model_cartoon} A depiction of the IB problem applied to a Gaussian field theory for $d = 2$, as described in Eq. \eqref{eq:hamiltonian_conditional}. Each column represents a different random variable ($\tilde{X}$, $X$, or $Y$) in the IB problem, while each row depicts a sample drawn from the joint distribution between them. Using $h$ as the relevance variable $Y$, the GIB-optimal coarsened field $\tilde{\chi}$ can be constructed through non-deterministic coarsening of $\chi$, as depicted by the arrows. The Lagrange multiplier $\beta_1$ controls the trade-off between minimizing mutual information between $\tilde{\chi}(\beta_1)$ and $\chi$ while maximizing mutual information between $\tilde{\chi}(\beta_1)$ and $h$. According to the semigroup structure described in Sec. \ref{sec:semigroup}, this process can be repeated to generate $\tilde{\chi}(\beta_2 \circ \beta_1)$ through non-deterministic mapping from $\tilde{\chi}(\beta_1)$ with compression level $\beta_2$. }
\end{figure*}


\subsubsection{IB regulator when disorder correlations are diagonal in momentum space}

In the first and simplest case, we take $H$ to be a $\delta$-function multiplied by some constant factor $\eta$. Since the Fourier transform $\mathcal{F}$ is unitary \footnote{We choose the formalism in which $[\mathcal{F}](k,x)$ includes a factor of $(2\pi)^{-d/2}$ so that $\mathcal{F}^\dagger\mathcal{F} = I$. Further, the appearance of factors $\delta^d(q-q')$ as opposed to the more traditional $\delta^d(q+q')$ is a consequence of our decision to conjugate with respect to $\mathcal{F}^\dagger\, \cdot\, \mathcal{F} $, instead of $\mathcal{F} \,\cdot\,\mathcal{F}$. This appeals more to the matrix multiplication shorthand.}, the momentum-space representation of $H$ is unchanged from its position-space representation:
\begin{eqnarray*}
    H(x_1,x_2) &=& \eta \delta^d(x_1-x_2)\,; \\
    \tilde{H}(q_1,q_2) &=& [\mathcal{F} H \mathcal{F}^\dagger](q_1,q_2) \\
    &=& \eta \delta^d(q_1 - q_2) \,.
\end{eqnarray*}

The first step in GIB analysis is constructing the canonical correlation matrix $\Sigma_X^{-1}\Sigma_{X|Y}$, where we have chosen $X \leftrightarrow \chi$ and $Y\leftrightarrow h$. After a calculation involving only Gaussian integral identities and our definition of $P[\chi, h]$, we obtain:
\begin{equation*}
    \Sigma_\chi^{-1}\Sigma_{\chi|h} = (I + H G_0)^{-1}
\end{equation*}

Next, we find the right eigenfunctions $V(x,u)$ and corresponding eigenvalues $\lambda(u)$ of the correlation. For our current construction of $H$,
\begin{eqnarray*}
    V(x,q) &=& \mathcal{F^\dagger}(x,q) = (2\pi)^{-d/2}e^{i q\cdot x} \\
    \lambda(q) &=& (1 + \eta \tilde{G}_0(q))^{-1}\,,
\end{eqnarray*}

Where $\tilde{G}_0(q) = 1/(t + q^2)$ is obtained after Fourier transform of $G_0$. To finally obtain the IB regulator in a familiar form, we would like to find a way to express it completely in terms of $q$, $k$, and the various other parameters introduced in this application. However, equation \eqref{eq:R_IB_def} gives us $R^{(\text{IB})}$ in terms of the bottleneck parameter $\beta$, which has not been defined yet in this application. 

The crucial insight is to note that $\beta$ serves essentially the same role as $k$ in the typical theory. To find the explicit map between the two, we use the fact that critical bottleneck values $\beta(q)$ are defined in terms of the canonical correlation eigenvalues $\lambda(q)$ through $\beta(q) = (1-\lambda(q))^{-1}$. In this model, the critical bottleneck values are
\begin{equation*}
    \beta(q) = \frac{1}{\eta \tilde{G}_0(q)} + 1\,.
\end{equation*}

Using this map, we can replace $\beta$ with $\beta(k)$, where $k$ is the usual momentum cutoff. Doing so, we find that the IB regulator can be neatly expressed in terms of the Litim regulator:
\begin{eqnarray}
    R^{(\text{IB})}_k(q) &=& \frac{t + q^2}{t + q^2 + \eta} (k^2 - q^2) \Theta(k^2 - q^2) \nonumber\\
    &=& \lambda(q) R^{(L)}_k(q)\,.
\end{eqnarray}

In particular, the limit $\eta \to 0$ gives $R^{(\text{IB})}\to R^{(\text{L})}$. It is interesting that the Litim regulator appears in this expression, since its derivation invokes optimality principles which are not obviously connected to information bottleneck. 

\subsubsection{Momentum-space IB regulator with dispersion in disorder correlations}

Without changing our decision to make $H$ diagonal in Fourier basis, we can also add $q$-dependence to $\eta$. In this case, the steps taken above are essentially unchanged, and we end up with a slightly different regulator:

\begin{eqnarray*}
    R_k^{(\IB)}(q) &=& \lambda (q) \lrp{\frac{\eta(q)}{\eta(k)}\tilde{G}_0^{-1}(k) - \tilde{G}_0^{-1}(q)} \times \\ 
    & & \Theta\lrp{\frac{\eta(q)}{\eta(k)}\tilde{G}_0^{-1}(k) - \tilde{G}_0^{-1}(q)}
\end{eqnarray*}

With some manipulations, one could optionally re-write this in terms of $(1-x)\Theta(1-x)$ in order to appeal to the Litim description once again.

A new feature appears in the regulator scheme when $\eta$ is given $q$-dependence. For extreme choices of $\eta$, the ordering of modes can actually be reversed. To see how this is possible, note that fundamentally it is the IB parameter $\beta$ which sets the cutoff, while the critical values $\beta(q)$ define the mapping to $q$. Therefore, by picking, e.g., $\eta(q) \sim \tilde{G}^{-2}_0(q)$, one achieves a $\beta(q)$ which monotonically decreases with respect to $q$, meaning longer wavelength modes (lower $q$) actually get integrated out \textit{before} shorter ones. However, this construction presents some pathologies and is hard to interpret in the truly continuous case, so we will not explore it further here.

\subsubsection{Explicit form of the IB regulator in a more general case} 

In the last section we assumed a form of $H$ which was diagonal in Fourier basis. This assumption led us to a regulator scheme which could be interpreted as a soft cutoff in momentum space. In this section we explore an example in which $H$ is no longer diagonal in Fourier basis:
\begin{equation*}
    H = \eta\, \mathcal{F}^\dagger \tilde{G}^{-1/2}_0 \mathcal{F}_\alpha \tilde{G}_0 \mathcal{F}_\alpha^\dagger \tilde{G}^{-1/2}_0 \mathcal{F}
\end{equation*}

Where $\mathcal{F}_\alpha$ is the fractional Fourier transform through angle $\alpha$ and $\eta$ is a constant. Under this definition, we can again compute $\Sigma_\chi$ and find the spectrum of $\Sigma_\chi^{-1}\Sigma_{\chi|h}$. This yields eigenfunctions analogous to the plane wave solutions in last section, but indexed by a new parameter $u$ which can neither be interpreted as position nor wavenumber:
\begin{eqnarray*}
    V^\dagger[\cdot](u) &=& \lrc{\tilde{G}_0^{1/2}\mathcal{F}_\alpha^\dagger \tilde{G}_0^{-1/2}\mathcal{F}}[\cdot](u)\\
    \lambda(u) &=& (1 + \eta \tilde{G}_0(u))^{-1}
\end{eqnarray*}

Here, the notation $[\cdot]$ indicates that $V^\dagger$ is best conceptualized as a functional parameterized by $u$, where for instance the collective modes of $\chi(x)$ would be given by $V^\dagger[\chi](u)$. Stated differently, the leftmost operator $\tilde{G}_0^{1/2}$ is evaluated at $u$, and the rightmost is a Fourier transform over the integrand $[\cdot]$. Unfortunately, this solution is only formal, and cannot be visualized in the same manner as plane waves. In a true field theory, even with the trivial Gaussian setup, both $H(x_1,x_2)$ and $V(x,u)$ are poorly behaved when written as functions of $x$. When written as an integral in $q$, $V$ diverges when $|q_{\text{max}}| \to \infty$, and is discontinuous in both $x$ and $u$. One way to conceptualize this is by comparison with $G_0^{-1}$, which includes $\nabla^2$ and thus cannot be written as elementary functions of $x$. After Fourier transform, we can replace the operator description with a simple function of the continuous variables $q$. Similarly, although we cannot express $H$ and $V$ as functions of $x$, the various operators we are interested in can be written simply in the non-orthogonal basis defined by $V$:
\begin{eqnarray}
    \lrs{V^{-1} H V^{-\dagger}}(u_1,u_2) &=& \eta \delta^d(u_1 - u_2) \\
    \lrs{V^\dagger G_0 V}(u_1,u_2) &=& \tilde{G}_0(u_1)\delta^d(u_1 - u_2)
\end{eqnarray}

It is hard to say what the label $u$ physically represents beyond being a parameter that defines and orders collective modes $\chi'(u) = V^\dagger [\chi](u)$ in the system. Despite this, the regulator maintains its simple form:
\begin{equation}
    R^{(\IB)}_v(u) = \lambda(u)R^{(\text{L})}_v(u)
\end{equation}

Where now $v$ takes the role of the cutoff, replacing $k$ as $u$ has replaced $q$. That is, the collective modes labelled by $u$ are ordered in terms of their predictiveness about the disordered source field $h$. GIB then imposes a soft-cutoff scheme at a scale $v$, which is a proxy for the bottleneck parameter $\beta$, as $k$ was in the Fourier case. We stress that these labels $v$ and $u$ are defined by the correlation structure of $P[\chi,h]$ and have no simple intrinsic physical meaning. Without significantly more effort, all we can say is that a mode labelled $u_1$ carries more information about the disorder $h$ than a mode labelled $u_2$ if $u_1 < u_2$.

Many of the difficulties present in this discussion, such as the poorly-behaved character of collective modes $V(x,u)$ and disorder correlator $H(x_1,x_2)$, as well as the non-intuitive nature of the mode labels $u$ and $v$, stem from a common cause. IB is only suited to analysis of systems with finitely many degrees of freedom, and field theories have infinitely many. The calculations above were nonetheless performed in this context to demonstrate that IB defines collective modes of a system and establishes a cutoff scheme which, in general, differs from traditional notions of relevance, as represented by the Fourier basis and momentum cutoff. This idea could be crucial to understanding collective behavior in systems without clear notions of locality or organization. Such problems abound in, for example, the brain where long-distance connections between brain areas are common and important for computation while information is also spread across many areas and recombined for important, multi-modal tasks. The recurrent, highly interconnected, and still computationally efficient structure in the brain renders the simple notion of physical distance between cells rather limiting.



\subsection{The relevance variable $Y$ can have many physical interpretations}

Gaussian IB begins with a choice of joint distribution $P(x,y)$. As we have discussed, this distribution gives a constrained parameterization of a cutoff scheme which is analogous to the one employed in Wetterich NPRG. In the last section, we showed that not all choices $P(x,y)$ lead to collective modes $V^T X$ which have a canonical interpretation such as Fourier modes. That discussion was carried out under the assumption that the relevance variable $Y$ pertains to a source field with some disorder statistics. Generally speaking, this is only one way of constructing $Y$. Even within the constraint of $P(x,y)$ being jointly Gaussian, the physical interpretation of $x$ and $y$ can vary. Here we briefly discuss some of these alternative interpretations.

First, $Y$ may represent the environment of a set of variables $X$. This scenario is analogous to the one presented by Koch-Janusz et al. \cite{koch-janusz_mutual_2018}. Consider a collection of spins on a lattice, and choose some enclosed region. Let $X$ be the state of the spins in that region and let $Y$ denote the state of those outside. In the case that these spins have Gaussian statistics, this is a valid starting point for GIB. With this setup, we expect that the most relevant collective modes would be relatively slowly varying in position. In fact, Gordon et al. recently formalized this idea for field theories not restricted to Gaussian statistics \cite{gordon_relevance_2021}. They consider a ``buffer'' zone between $X$ and $Y$ whose size is taken to infinity. In this limit, the first collective variables encoded by IB at strong compression (low $\beta$ in our notation) correspond to the operators with the smallest scaling dimensions, and hence the most relevant operators in the RG sense. Their approach is therefore promising for the analysis of systems with local interactions whose order parameter is not known \textit{a priori}. More fundamentally, they have shown that $Y$ and $X$ can be chosen to enforce a traditional, ``physical'' definition of relevance. 

Second, consider a stationary stochastic process with Gaussian statistics both in time and across variable index. We could choose $X$ to represent the current state of the system while $Y$ represents the future. Here, the most relevant modes are those projections of $X$ which vary the slowest. In fact, if we suppose that time has been properly discretized, this interpretation of the GIB problem is equivalent to a certain class of slow feature analysis problems \cite{creutzig_predictive_2008}. 

Third, we can imagine another dynamical system in which variables $X$ which are driven by a stochastic signal $Y$ such that the joint distribution is Gaussian and stationary. Now, the features of $X$ which are most relevant are no longer simply the slowest-varying components. The cutoff scheme we find will depend on the statistics which generate $Y$, the manner in which $Y$ couples to $X$, the internal dynamics of $X$, and whether we take $Y$ to be in the past, future, or present. 

Together with the example from last section, in which $Y$ fulfilled the role of a disordered source field, these examples span a number of physically interesting scenarios. Certainly, more are possible. Any valid interpretation will generally consist of a set of random variables $\{Z_i\}$ that obeys a Gaussian joint distribution, which is then partitioned into two or three disjoint sets. The first is $\{X_n\}$, the second is $\{Y_m\}$, and the third, which is optional, is a dummy set containing every $Z_i$ which we don't care to include in the model. In the case that these sets aren't disjoint, it is possible to have $X$ and $Y$ become deterministically related which is an invalid starting point for GIB. Finally, we note that while this framework allows for some discussion of systems involving dynamics, it is poorly suited for application to general stochastic processes as the distribution $P(X,Y)$ must be stationary. This also means that the connections drawn here between GIB and NPRG are \textit{not} meant to cover the more general, dynamical NPRG framework often seen in nonequilibrium statistical mechanics literature \cite{kopietz_introduction_2010,canet_general_2011,haga_renormalization_2019}. However, given the importance of both IB and the dynamical NPRG to applications in nonequilibrium settings, we believe that a more general framework is in demand.

\section{Conclusion}

In this manuscript, we have examined structural similarities between the Gaussian information bottleneck problem and a class of RG techniques involving soft cutoffs. Our main result is to identify that the crucial connection between the two is a non-deterministic coarsening map. In NPRG, this map defines both the UV-regulated coarse-grained Hamiltonian of the Wilson-Polchinski picture, as well as the IR-regulated free energy used in the Wetterich approach. Therefore, one can rigorously connect IB to RG by requiring that this coarsening map solves a particular IB problem. In doing so, one parameterizes a space of soft cutoff schemes in terms of IB relevance variable statistics $P(x,y)$. Additionally, one can identify the structures in an IB problem which are analogous to UV- and IR-cutoffs in RG. 

While we believe that this connection holds for more general IB problems, we limited our discussion to Gaussian statistics for two main reasons. First, NPRG coarsening maps are always Gaussian, since this leads to simpler flow equations with physical interpretations. Second, in order to be compatible with this first consideration, we studied only the GIB problem which has exactly known solutions that are Gaussian \cite{chechik_information_2005}. 

Another result was to show that the GIB coarsening map satisfies a semigroup property. In particular, we identify an explicit function $b(\beta)$ which multiplies under composition of coarsening maps in a manner analogous to the length scale in a traditional RG setting. Given that the typical role of semigroup structure in RG theory is the identification of anomalous exponents, it is not within the scope of this manuscript to assign a similar task to $b(\beta)$. More immediately, the presence of this structure within GIB raises the question of whether it may be present in IB schemes more generally. If so, would an iterative coarse-graining scheme consisting of repeated low-compression transformations be advantageous as an analysis technique?

By explicitly comparing the set of GIB solutions provided by Chechik et al. with a generic NPRG scheme, we identified the IR cutoff scheme present in GIB \eqref{eq:R_IB_def}. A similar analysis can be carried out to identify the UV cutoff, but doing so involves a discussion about reparameterization which we felt would distract from the main points. Direct computations on a toy model showed that the IB regulator has some characteristics which are similar to the ubiquitous Litim regulator \cite{litim_optimized_2001}. An important generalization is that IB selects the collective mode basis according to which features of the system state $X$ will be most informative about $Y$, whatever it is chosen to be. We gave a simple example in which this collective mode basis could not be interpreted as a Fourier basis. In general, this will be the case, though depending on how $Y$ is defined, one may still arrive at collective modes which are essentially Fourier in nature. One bit of analysis we did not carry out is the connection of IB to the dynamical NPRG, though for non-equilibrium problems involving IB---such as the predictive coding problem---this may be a fruitful avenue for further work.

Next, we note that IB is generally extremely difficult to solve, so restricting an NPRG scheme to a family of exact IB solutions is completely unrealistic without significant advances in IB theory. One avenue of attack is to find better ways of solving IB. As outlined in sec. \ref{sec:BA}, a more general parametric Blahut-Arimoto scheme would be very powerful in this context since it could essentially replace the flow-equation description with a self-consistency scheme at each cutoff value. However, given that the exact Gaussian form we derive is complicated, this seems unlikely to work. A more realistic approach to practical IB-RG implementation is to relax the IB-optimality constraint. We suggest that even in a non-Gaussian setting, one could directly calculate the IB regulator \eqref{eq:R_IB_def} proposed here and use the NPRG flow equations in exactly the same way. While the resulting statistics would no longer be exactly IB-optimal, this procedure is no more difficult than any other NPRG implementation, and may produce qualitatively similar results to an exact IB solution.

We reiterate that not all IB problems will benefit from the RG connections presented here, and vice versa. Ideally, the problem in question involves a system with a large, but finite, number of degrees of freedom $X$ statistically coupled to a similarly large number of random variables $Y$. Finiteness is required by IB, but because of the construction of the NPRG, this is not an issue. The flow is defined exactly even in the absence of a traditional rescaling step, which would be illegal in a finite system since it adds more modes. Biophysics systems, for example, may be particularly well-suited to IB-RG analysis, because $Y$ can be chosen to have biological relevance, and the cutoff scheme will define and prioritize collective modes that are most informative about that function. Biological systems all have size and energy constraints that make the efficient compression of inputs from the external world critical for survival. Balancing that, and just as important for function, organisms also have clear preference for what is relevant in that external signal, namely which aspects can be used to drive behavior that confers a fitness benefit. The IB framework helps cast behavioral relevance as the prime mover in input compression, while the RG can help show how this kind of computation is achieved. Uniting these theories can provide a way to pull together normative notions of relevance with their mechanistic implementation.

\begin{acknowledgments}

We thank Umang Mehta and David J.\ Schwab for comments. This work was supported by the National Institutes of Health BRAIN initiative program (R01B026943) and by the Aspen Center for Physics, which is supported through National Science Foundation grant \mbox{PHY-1607611}.

\end{acknowledgments}

\appendix

\section{\label{app:semigroup}Detailed derivation of GIB semigroup structure}
A map $(A,\Sigma_\xi)$ representing $\tilde{X} = A X + \xi$ solves the GIB problem if it satisfies:
\begin{eqnarray} \label{eq:GIB_opt_cond}
    [V^{-1} A^T & & \Sigma_\xi^{-1}A V^{-T}]_{ij} = \\
    & & \frac{\beta(1-\lambda_i)-1}{s_i\lambda_i} \Theta\lrp{\beta - \frac{1}{1-\lambda_i}}\delta_{ij} \nonumber
\end{eqnarray}
\noindent
for some $\beta$. To show that the composition of two GIB maps is IB-optimal, we explicitly compute the above expression for the map $(A,\Sigma_\xi)$ arrived at by sequential coarsening. The individual maps are,
\begin{eqnarray*}
   \tilde{X}_1 &=& A_1 X + \xi_1\\
   \tilde{X}_2 &=& A_2 \tilde{X}_1 + \xi_2\,.
\end{eqnarray*}
\noindent
This construction gives
\begin{eqnarray*}
    \tilde{X}_2 &=& A_2 A_1 X + A_2 \xi_1 + \xi_2\\
        &=& A X + \xi
\end{eqnarray*}
\noindent
So we have that, for $\Sigma_{\xi_1} = \Sigma_{\xi_2} = I$,
\begin{equation*}
    (A,\Sigma_\xi) = (A_2A_1, A_2 A_2^T + I)\,.
\end{equation*}

In order to ensure that both $A_1$ and $A_2$ are diagonal, we project $X$ into natural basis with the replacement $X \to V^T X$. Note that $A_2$ is actually automatically diagonal because the first compressed representation $\tilde{X}_1 = A_1 X + \xi_1$ is already in natural basis. After this transformation, the optimality condition \eqref{eq:GIB_opt_cond} is simplified because the $V^{-1}$ matrices have been absorbed into the definition of $X$. The new condition is:
\begin{eqnarray} \label{eq:GIB_opt_cond_simplified}
    [A^T &&\Sigma_\xi^{-1}A]_{ij} = \\ &&\frac{\beta(1-\lambda_i)-1}{s_i\lambda_i} \Theta\lrp{\beta - \frac{1}{1-\lambda_i}}\delta_{ij} \nonumber
\end{eqnarray}

Now we explicitly compute $A_1$ and $A_2$. From \eqref{eq:GIB_sol} we have:
\begin{eqnarray*}
    [A_{1}]_{ij} &=& \lrs{\frac{\beta_1 (1-\lambda_i) - 1}{s_i\lambda_i}}^{1/2}\Theta\lrp{\beta_1 - \frac{1}{1-\lambda_i}}\delta_{ij} \\~ 
    [A_{2}]_{ij} &=& \lrs{\frac{\beta_2 (1-\lambda'_i) - 1}{s'_i\lambda'_i}}^{1/2}\Theta\lrp{\beta_2 - \frac{1}{1-\lambda'_i}}\delta_{ij}
\end{eqnarray*}
\noindent
where
\begin{eqnarray*}
    [\Sigma_{X|Y}]_{ij} &=& s_i \lambda_i \delta_{ij} \\~
    [\Sigma_X]_{ij} &=& s_i\delta_{ij} \\~
    [\Sigma_{\tilde{X}_1|Y}]_{ij} &=& s'_i \lambda'_i \delta_{ij} \\~
    [\Sigma_{\tilde{X}_1}]_{ij} &=& s'_i\delta_{ij}
\end{eqnarray*}

The latter two equations must be re-expressed in terms of the original $X-Y$ statistics, represented by $\lambda_i$ and $s_i$.
\begin{eqnarray*}
    \Sigma_{\tilde{X}_1 | Y} &=& A_1 \Sigma_{X|Y} A_1^T + I\\
    &\Rightarrow& s'_i\lambda'_i = \lambda_i s_i [A_1]^2_{ii} + 1 \\
    \Sigma_{\tilde{X}_1} &=& A_1 \Sigma_X A_1^T + I \\~ &\Rightarrow& s'_i = s_i [A_1]^2_{ii} + 1
\end{eqnarray*}

Solving for $\lambda'$, we have:
\begin{equation*}
\lambda'_i = \frac{\lambda_i s_i [A_1]^2_{ii} + 1}{s_i [A_1]^2_{ii}+1}
\end{equation*}
Now, directly evaluating $A_1$ and $s_i$,  we get the following for $\lambda'$ and $s'$:
\begin{eqnarray*}
    \lambda'_i &=& \min\lrc{\frac{\beta_1}{\beta_1 - 1} \lambda_i,\, 1} \\
    s'_i &=& \frac{\beta_1(1 - \lambda_i)- 1}{\lambda_i}\Theta\lrp{\beta_1 - \frac{1}{1 - \lambda_i}} + 1
\end{eqnarray*}
\begin{widetext}
Using these last two expressions, $A_2$ can be expressed directly in terms of $s$ and $\lambda$. By direct substitution, we can now check whether the composite scheme $(A,\Sigma_\xi)$ satisfies the GIB optimality condition \eqref{eq:GIB_opt_cond_simplified}:
\begin{eqnarray*}
[A\Sigma_\xi^{-1}A^T]_{ij} &=& [A_2A_1(A_2^2 + I)^{-1}A_1A_2]_{ij} \\~
&=& \frac{(\beta_1(1-\lambda_i) - 1)(\beta_2(1-\frac{\beta_1}{\beta_1-1}\lambda_i)-1)}{s_i\lambda_i(\beta_1(1-\lambda_i) + \beta_2(1-\frac{\beta_1}{\beta_1-1}\lambda_i)-1)} \Theta\lrp{\beta_2 - \frac{1}{1-\min\lrc{1,\,\frac{\beta_1}{\beta_1 -1}\lambda_i}}} \delta_{ij}\\
&=& \frac{(\beta_2 \circ \beta_1)(1-\lambda_i) - 1}{s_i\lambda_i} \Theta\lrp{\beta_2 \circ \beta_1 - \frac{1}{1-\lambda_i}} \delta_{ij}
\end{eqnarray*}
\end{widetext}
\noindent
where the binary operator $\circ$ is given by
\begin{equation*}
    \beta_2 \circ \beta_1 =  \frac{\beta_2 \beta_1}{\beta_2 + \beta_1 - 1} \,.
\end{equation*}

By identifying $\beta_2 \circ \beta_1$ with a single value $\beta$, we find that the GIB optimality condition \eqref{eq:GIB_opt_cond} is satisfied. It is important to note that this operator maps the space of valid $\beta$ values $\mathbb{R}>1$ to itself. That is,
\begin{equation*}
    \circ : \mathbb{R}>1\, \times\, \mathbb{R}>1\, \to\, \mathbb{R}>1
\end{equation*}
\noindent
Which means that $\beta_2 \circ \beta_1$ really can be identified as a bottleneck parameter. Along with associativity, this means that $(\mathbb{R} >1,\circ)$ is a semigroup representing sequential GIB coarsening.

\section{Derivation of the GIB regulator}

In section \ref{sec:GIB_regulator} we present an IR regulator that is both analogous to the one used in the Wetterich NPRG formalism, and which enforces optimality in the Gaussian IB problem. Here, we show explicitly how this regulator is derived. To begin, recall that the role of the regulator is to deform the microscopic theory through the addition of a mass-like term which ``freezes out'' the most relevant modes:
\begin{equation*}
    \Delta S_k[\chi] = \frac{1}{2} \chi^\dagger R_k \chi
\end{equation*}

In a context relevant to GIB where the bare variable $x$ is finite-dimensional, we would write this as:
\begin{equation*}
    \Delta S_\beta(x) = \frac{1}{2} x^\dagger R_\beta x
\end{equation*}
\noindent
With $R_\beta$ a positive semi-definite matrix. Following the argument in section \ref{sec:soft_is_nondeterminsitic}, we can identify the deformation produced by a Gaussian coarsening $\tilde{X} = A X + \xi$:
\begin{equation*}
    \Delta S(x) = \frac{1}{2} x^T A^T \Sigma_\xi^{-1} A x
\end{equation*}

Now, by imposing IB optimality \eqref{eq:GIB_opt_cond} on $(A,\Sigma_\xi)$, we find that 
\begin{eqnarray*}
    R_\beta^{(\IB)} &=& A(\beta)^T \Sigma_\xi(\beta)^{-1}A(\beta) \\
        &=& V\ \text{diag}(\alpha_i^2(\beta))\ V^T
\end{eqnarray*}

With 
\begin{equation*}
    \alpha^2_i(\beta) = \frac{\beta(1-\lambda_i) - 1}{\lambda_i s_i}\Theta\lrp{\beta - \frac{1}{1-\lambda_i}}\,.
\end{equation*}

After the substitution $\beta_i = (1-\lambda_i)^{-1}$,
\begin{equation*}
     \lrs{R^{(\IB)}_\beta}_{ij} = \sum_{u}V_{iu}\frac{\beta - \beta_u}{s_u (\beta_u - 1)} \Theta(\beta - \beta_u) V_{uj}
\end{equation*}
This expression differs from the one given in section \ref{sec:GIB_regulator} because there we assumed that $X$ had already been projected into natural basis. Here, taking $X \to V^TX$ means $R \to V^{-1} R V^{-T} $, and so
\begin{equation*}
    \lrs{R^{(\IB)}_\beta}_{ij} = \frac{\beta - \beta_i}{s_i (\beta_i - 1)} \Theta(\beta - \beta_i)\delta_{ij} \,.
\end{equation*}

\begin{widetext}

\section{\label{app:BA_updates}Blahut-Arimoto update scheme for GIB in terms of NPRG objects}

Eqs. \eqref{eq:BA_update} depict the Blahut-Arimoto updates for $\Sigma_{X|\tilde{X}}$, $\Sigma_{\tilde{X}}$, and $\Sigma_{X\tilde{X}}$ at a schematic level. Written as expectations, these matrices are:

\begin{eqnarray*}
    [\Sigma_{X|\tilde{X}}]_{ab} &=& \mathbb{E}_{X|\tilde{X}=\tilde{x}}\lrc{(X - \mu_{X|\tilde{X}}(\tilde{x}))_a (X - \mu_{X|\tilde{X}}(\tilde{x}))_b} \\~
    [\Sigma_{\tilde{X}}]_{ab} &=& \mathbb{E}_{\tilde{X}} \lrc{(\tilde{X} -\mu_{\tilde{X}})_a(\tilde{X} -\mu_{\tilde{X}})_b} \\~
    [\Sigma_{X\tilde{X}}]_{ab} &=& \mathbb{E}_{X,\tilde{X}}\lrc{(X - \mu_X)_a(\tilde{X}-\mu_{\tilde{X}})_b}
\end{eqnarray*}

As described in the main text, the BA procedure can be thought of as an iterative procedure wherein an estimate for $P(\tilde{x}|x)$, or equivalently $P(x,\tilde{x})$, is plugged into a known functional representing the consistency condition required by optimality. Schematically, 
\begin{eqnarray*}
    P'(x,\tilde{x}) &=& \text{BA}\lrs{P(x,\tilde{x})} \\~
        &=& \frac{1}{Z_\beta(x)}P(x)P(\tilde{x})\exp\lrs{-\beta  D_{\text{KL}}\lrs{P(y|x)||P(y|\tilde{x})}}
\end{eqnarray*}
\noindent
where $D_{\text{KL}}$ is the Kullback-Leibler divergence, defined for two distributions $P$ and $Q$ of the same variable as
\begin{equation*}
    D_{\text{KL}}\lrs{P||Q} = \int \dd y\, P(y) \log \frac{P(y)}{Q(y)}
\end{equation*}
\noindent
and the RHS can be seen as a functional of $P(x,t)$ through the expressions:
\begin{eqnarray*}
    P(\tilde{x}) &=& \int \dd x\, P(x,\tilde{x}) \\~
    P(y|\tilde{x}) &=& \frac{1}{P(\tilde{x})}\int \dd x\, P(y|x) P(x,\tilde{x}) \\
    Z_\beta(x) &=& \int \dd \tilde{x}\, P(\tilde{x})\exp\lrs{-\beta D_{\text{KL}}\lrs{P(y|x)||P(y|\tilde{x})}}\,.
\end{eqnarray*}

In this appendix, we derive the equations \eqref{eq:BA_update} using the explicit form of the BA map presented above. The goal is to express ``updates'' for the matrices $\Sigma_{X|\tilde{X}}$, $\Sigma_{\tilde{X}}$, and $\Sigma_{X\tilde{X}}$ in terms of their current estimates. In general, quantities describing the updated joint distribution $P'(x,\tilde{x})$ will be primed. To begin, we evaluate $P(y|\tilde{x})$ using elementary properties of Gaussian variables. Next, we evaluate the divergence $D_\text{KL}$, and the partition function $Z_\beta(x)$. Finally, we combine these elements and read off the updated parameters.
\noindent
Suppose $a = b + c$, with $b \sim \mathcal{N}(\mu_b,\Sigma_b)$ and $c\sim\mathcal{N}(\mu_c, \Sigma_c)$. Then
\begin{equation*}
    a \sim \mathcal{N}(\mu_a, \Sigma_a) \quad \text{with} \quad \mu_a = \mu_b + \mu_c \,,\quad \Sigma_a = \Sigma_b + \Sigma_c
\end{equation*}
\noindent
Therefore consider $y = Wx + z$ with $z \sim \mathcal{N}(0, \Sigma_{Y|X})$ and suppose that $P(x|\tilde{x}) = \mathcal{N}(\mu_{X|\tilde{X}}, \Sigma_{X|\tilde{X}})$. Then
\begin{align*}
    \mu_{Y|\tilde{X}} &= W \mu_{X|\tilde{X}} \\~
    \Sigma_{Y|\tilde{X}} &= W \Sigma_{X|\tilde{X}} W^T + \Sigma_{Y|X}
\end{align*}
\noindent
Now, consider jointly Gaussian variables $(a,b)$. Then
\begin{equation*}
    \mu_{a|b} = \mu_a + \Sigma_{ab}\Sigma_b^{-1}(b - \mu_b)
\end{equation*}
\noindent
Hence, assuming without loss of generality that $\mu_Y=0$, $\mu_X = 0$, and $\mu_{\tilde{X}} = 0$, 
\begin{equation*}
    \mu_{Y|X} = W x \quad \Rightarrow \quad W = \Sigma_{YX}\Sigma_{X}^{-1} 
\end{equation*}
\noindent
and
\begin{equation*}
    \mu_{X|\tilde{X}} = \Sigma_{X\tilde{X}}\Sigma_{\tilde{X}}^{-1}\tilde{x}
\end{equation*}
\noindent
so finally,
\begin{eqnarray*}
    \Sigma_{Y\tilde{X}} &=& \Sigma_{YX}\Sigma_X^{-1} \Sigma_{X\tilde{X}} \\~
    \Sigma_{Y|\tilde{X}} &=& \Sigma_{YX} \Sigma_{X}^{-1}\Sigma_{X|\tilde{X}} \Sigma_{X}^{-1}\Sigma_{XY} + \Sigma_{Y|X} 
\end{eqnarray*}

These matrices allow us to construct $P(y|\tilde{x})$ and thereby calculate $D_{\text{KL}}$. For Gaussian distributions, the KL divergence has a standard form. In this context, we care only about the terms which carry $x$ and $\tilde{x}$ dependence.
\begin{eqnarray*}
    D_{\text{KL}}[P(y|x)||P(y|\tilde{x})]  &\sim&  \frac{1}{2}(\mu_{Y|X} - \mu_{Y|\tilde{X}})^T\Sigma_{Y|\tilde{X}}^{-1}(\mu_{Y|X} - \mu_{Y|\tilde{X}}) \\~
        &=& \frac{1}{2} \tilde{x}^T \Sigma_{\tilde{X}}^{-1}\Sigma_{\tilde{X}Y} \Sigma_{Y|\tilde{X}}^{-1} \Sigma_{Y\tilde{X}} \Sigma_{\tilde{X}}^{-1}\tilde{x} -  x^T \Sigma_{X}^{-1}\Sigma_{XY}\Sigma_{Y|T}^{-1}\Sigma_{Y\tilde{X}}\Sigma_{\tilde{X}}^{-1} \tilde{x} + \lrc{x^2} \\~
        &=& \frac{1}{2} \tilde{x}^T \Sigma_{\tilde{X}}^{-1}\Sigma_{\tilde{X}Y} \Sigma_{Y|\tilde{X}}^{-1} \Sigma_{Y\tilde{X}} \Sigma_{\tilde{X}}^{-1}\tilde{x} -  x^T B^T \tilde{x} + \lrc{x^2}
\end{eqnarray*}

Where $\sim$ denotes ``up to addition of a constant.'' The matrix $B$ describing the coupling between $x$ and $\tilde{x}$ has been introduced for convenience. Note also that there is a pure-$x$ term in this quantity, denoted $\lrc{x^2}$, which will cancel with the partition function $Z_\beta(x)$ that normalizes $P(t|x)$ in the BA map. In addition to this trivial $x$-dependence, $Z_\beta(x)$ also contributes a new $x^2$ term, which needs to be included. 
\begin{eqnarray*}
Z_\beta(x) &=& \int \dd \tilde{x}\, P(\tilde{x}) \exp\lrp{-\beta D_{\text{KL}}\lrs{P(y|x)||P(y|\tilde{x})}} \\~
&=& \int \dd \tilde{x}\, \exp\lrp{-\frac{1}{2}\tilde{x}^T\lrs{\Sigma_{\tilde{X}}^{-1} + \beta \Sigma_{\tilde{X}}^{-1}\Sigma_{\tilde{X}Y}\Sigma_{Y|\tilde{X}}^{-1}\Sigma_{Y\tilde{X}}\Sigma_{\tilde{X}}^{-1}}\tilde{x} + \beta x B^T\tilde{x} + \lrc{x^2} + \text{consts.}} \\~
&=& \int \dd \tilde{x}\, \exp\lrp{-\frac{1}{2}\tilde{x}^T\Sigma'^{-1}_{\tilde{X}|X}\tilde{x} + \beta x^T B^T\tilde{x} + \lrc{x^2} + \text{consts.}} \\~
&\sim& \exp\lrp{\frac{1}{2}\beta^2 x^T B^T \Sigma'_{\tilde{X}|X}B x + \lrc{x^2}}
\end{eqnarray*}

Here we have introduced $\Sigma'_{\tilde{X|X}}$ to further clean up notation. Now it is straightforward to obtain $P'(x,t)$ from the BA map by direct evaluation.
\begin{eqnarray*}
    P'(x,t) &=& Z_\beta(x)^{-1}P(x)P(\tilde{x})\exp\lrs{-\beta  D_{\text{KL}}\lrs{P(y|x)||P(y|\tilde{x})}} \\~
    &=& \exp\lrp{-\frac{1}{2}x^T\Sigma_{X}^{-1}x - \frac{1}{2}\beta^2 x^T B^T \Sigma'_{\tilde{X}|X} B x -\frac{1}{2} \tilde{x}^T \Sigma'^{-1}_{\tilde{X}|X}\tilde{x} + \beta x^T B^T \tilde{x}}
\end{eqnarray*}
Now, finally, all that remains is to complete the square and put the distribution in the form:
\begin{equation*}
    P'(x,\tilde{x}) \sim \exp\lrp{-\frac{1}{2}(x - \mu'_{X|\tilde{X}})^T\Sigma'^{-1}_{X|\tilde{X}}(x - \mu'_{X|\tilde{X}}) - \frac{1}{2}\tilde{x}^T\Sigma'^{-1}_{\tilde{X}}\tilde{x}}
\end{equation*}
\noindent
where
\begin{equation*}
    \mu_{X|\tilde{X}} = \Sigma'_{X\tilde{X}}\Sigma'^{-1}_{\tilde{X}} \tilde{x}
\end{equation*}
\noindent
After completing the square, the updated matrices $\Sigma'_{X|\tilde{X}}$, $\Sigma'_{\tilde{X}}$, and $\Sigma'_{X\tilde{X}}$ can be read off:
\begin{eqnarray*}
    \Sigma'^{-1}_{X|\tilde{X}} &=& \Sigma_{X} + \beta^2 B^T\Sigma'_{\tilde{X}|X}B \\~
    \Sigma'^{-1}_{\tilde{X}} &=& \Sigma'^{-1}_{\tilde{X}|X} - \beta^2 B\Sigma'_{X|\tilde{X}} B^T \\~
    \Sigma'_{X\tilde{X}} &=& \beta \Sigma'_{X|\tilde{X}} B^T \Sigma'_{\tilde{X}} \,.
\end{eqnarray*}
These can be written entirely in terms of the old estimates through the substitutions:
\begin{eqnarray*}
    B^T &=& \Sigma_{X}^{-1}\Sigma_{XY}\Sigma_{Y|T}^{-1}\Sigma_{Y\tilde{X}}\Sigma_{\tilde{X}}^{-1} \\~
    \Sigma'^{-1}_{\tilde{X}|X} &=& \Sigma_{\tilde{X}}^{-1} + \beta \Sigma_{\tilde{X}}^{-1}\Sigma_{\tilde{X}Y}\Sigma_{Y|\tilde{X}}^{-1}\Sigma_{Y\tilde{X}}\Sigma_{\tilde{X}}^{-1} \\~
    \Sigma_{Y\tilde{X}} &=& \Sigma_{YX}\Sigma^{-1}_{X}\Sigma_{X\tilde{X}} \\~
    \Sigma_{Y|\tilde{X}} &=& \Sigma_{Y|X} + \Sigma_{YX} \Sigma^{-1}_{X}\Sigma_{X|\tilde{X}}\Sigma^{-1}_{X}\Sigma_{XY}
\end{eqnarray*}

Finally, we note that iteration of these equations does not guarantee the convergence of each matrix involved, since invertible linear transformations on the random variables are a symmetry of the objective function. The GIB-optimal solutions, which are described by the fixed points of this update scheme, are connected continuously by these symmetries. If one wishes to use these updates practically and ensure that all matrices converge to fixed values, it is necessary to break this reparameterization invariance by taking extra steps after each update. In the original GIB paper \cite{chechik_information_2005}, the reparameterization-invariant quantities $\alpha_i$ are instead plotted over iteration of their BA scheme, because their convergence is guaranteed. 

\section{\label{app:field_theory}Selected computations for toy model}
\subsection{Canonical correlation Green's function}
A central object in GIB is the canonical correlation matrix, $\Sigma_{X}^{-1}\Sigma_{X|Y}$. From this object, one obtains the eigenvector matrix $V$, which describes the linear transformation of $X$ into its collective modes, and eigenvalues $\lambda_i$, which order these modes in terms of their information content about $Y$. In the toy model, we begin with physical definitions for the statistics in Eqs. \eqref{eq:hamiltonian_conditional} and \eqref{eq:disorder_avg}. Then, by interpreting $\chi$ as the input variable $X$ and the disorder $h$ as the relevance variable $Y$, we ask what the structure of the resulting GIB-regularized NPRG scheme looks like. Like any other GIB problem, we must first calculate the canonical correlation Green's function, $\Sigma_\chi^{-1}\Sigma_{\chi|h}$. Two Green's functions come directly from the definitions:
\begin{equation*}
   \Sigma_{\chi|h} = G_0\,, \qquad \Sigma_h = H
\end{equation*}
\noindent
To find $\Sigma_{\chi}$, we need $\Sigma_{\chi h}$, which we get through $\mu_{\chi|h}$:
\begin{equation*}
    \mu_{\chi|h} = \Sigma_{\chi h}\Sigma_{h}^{-1}h
\end{equation*}
\noindent
Compute this mean by looking at the Hamiltonian for $\chi$ with frozen disorder $h$:
\begin{eqnarray*}
\mathcal{H}[\chi|h] &=& \frac{1}{2} \chi^T \Sigma_{\chi|h}^{-1}\chi - h^T \chi \\ 
&=& \frac{1}{2}(\chi - \mu_{\chi|h})^T\Sigma_{\chi|h}^{-1}(\chi - \mu_{\chi|h}) - \frac{1}{2} \mu_{\chi|h}^T\Sigma_{\chi|h}^{-1}\mu_{\chi|h} \\ 
\Rightarrow h &=& \Sigma_{\chi|h}^{-1}\mu_{\chi|h} \\ 
\end{eqnarray*}
\noindent
Hence we can identify $\Sigma_{\chi h}$:
\begin{equation*}
    \Sigma_{\chi h} = \Sigma_{\chi|h}\Sigma_{h} = G_0 H
\end{equation*}
\noindent
Now, use the Schur complement formula to identify $\Sigma_\chi$:
\begin{eqnarray*}
    \Sigma_\chi &=& \Sigma_{\chi|h} + \Sigma_{\chi h}\Sigma_{h}^{-1} \Sigma_{h\chi} \\
     &=& G_0 + G_0 H H^{-1} H G_0 \\
     &=& G_0 + G_0 H G_0
\end{eqnarray*}
So finally, the canonical correlation Green's function is 
\begin{equation*}
    \Sigma_{\chi}^{-1}\Sigma_{\chi|h} = (I + H G_0)^{-1}
\end{equation*}

\subsection{Canonical correlation eigendecomposition calculations}
\subsubsection{Fourier collective basis}
Once the canonical correlation Green's function is known, one calculates its eigenfunctions (or eigenvectors, in the usual, finite-dimensional case) and eigenvalues. In the main text, we consider three constructions of $H$ which altogether yield two eigenbases: Fourier and non-Fourier. Let's first calculate the spectrum $\lambda(q)$ for case \textit{2} in section \ref{sec:not_always_fourier}, which also covers the analysis of case \textit{1}. 
\begin{eqnarray*}
    H(x_1, x_2) &=& \frac{1}{(2\pi)^d} \int \dd^dq\, \eta(q) e^{iq\cdot(x_1 - x_2)} \\
    &=& [\mathcal{F}^\dagger \tilde{H} \mathcal{F}](x_1,x_2)
\end{eqnarray*}

Where $\tilde{H}$ represents a ``diagonal'' function, $\tilde{H}(q_1,q_2) = \eta(q_1)\delta^d(q_1 - q_2)$. The frozen disorder propagator $G_0$ is also diagonal in Fourier basis:
\begin{eqnarray*}
    G_0(x_1,x_2) &=& \delta^d(x_1-x_2)[t - \nabla_{x_2}^2]^{-1} \\ 
    &=& \frac{1}{(2\pi)^d}\int \dd^d q\,\frac{1}{t + q^2} e^{iq\cdot(x_1 - x_2)} \\
    &=& [\mathcal{F}^\dagger \tilde{G}_0\mathcal{F}](x_1,x_2)
\end{eqnarray*}

We use $\tilde{G}_0(q)$ to represent both the function $(t + q^2)^{-1}$, and the diagonal kernel $(t+q^2)^{-1}\delta^d(q_1-q_2)$ interchangeably, as needed. Using the expression for $\Sigma_{\chi}^{-1}\Sigma_{\chi|h}$ derived in the last section, we have:
\begin{eqnarray*}
    \Sigma_{\chi}^{-1}\Sigma_{\chi|h} &=& (I + H G_0)^{-1} \\
    &=& (I + \mathcal{F}^{\dagger}\tilde{H}\mathcal{F} \mathcal{F}^\dagger\tilde{G}_0\mathcal{F})^{-1}\\
    &=&\mathcal{F}^\dagger (I + \tilde{H}\tilde{G}_0)^{-1}\mathcal{F} \\
    &=& V \Lambda V^{-1}
\end{eqnarray*}
Since $\mathcal{F}$ is unitary and both $\tilde{H}$ and $\tilde{G}_0$ are diagonal, we have:
\begin{equation*}
    V(x,q) = \mathcal{F}^\dagger(x,q) = \frac{1}{(2\pi)^{d/2}}e^{iq\cdot x}\,, \qquad \lambda(q) = \frac{1}{1 + \eta(q)\tilde{G}_0(q)}
\end{equation*}

\subsubsection{Non-Fourier collective basis}
Next, we carry out the same computation for case \textit{3}, in which $H$ is not diagonal in Fourier basis, and so neither is the canonical correlation Green's function. Written formally, the disorder correlator is
\begin{equation*}
    H = \eta \mathcal{F}^\dagger \tilde{G}_0^{-1/2}\mathcal{F}_\alpha \tilde{G}_0 \mathcal{F}_\alpha^\dagger \tilde{G}_0^{-1/2}\mathcal{F}
\end{equation*}
\noindent
Where $\mathcal{F}_\alpha$ is the $d$-dimensional fractional Fourier transform. The 1-dimensional version defined as:

\begin{equation} \label{eq:frac_fourier}
    \mathcal{F}^{(1)}_\alpha[f](u) = (2\pi i \sin \alpha)^{-1/2} \int_{\mathbb{R}} \dd x\, f(x) \exp \lrs{- i \lrp{\csc(\alpha) u x - \frac{1}{2}\cot(\alpha)(x^2 + u^2)}} 
\end{equation}

This transform is unitary, satisfies $\mathcal{F}^{(1)}_\alpha = \mathcal{F}^{(1)\dagger}_{-\alpha}$, and $\alpha = \pi/2$ gives the usual one-dimensional Fourier transform. To construct the $d$-dimensional version $\mathcal{F}_\alpha$, we simply take tensor products: $\mathcal{F}_\alpha$ = $\mathcal{F}^{(1)}_\alpha \otimes \cdots \otimes \mathcal{F}^{(1)}_\alpha$ with $d$ copies. Hence, $\mathcal{F}_\alpha$ has properties analogous to $\mathcal{F}^{(1)}_\alpha$, namely
\begin{equation*}
    \mathcal{F}_\alpha^\dagger = \mathcal{F}_{-\alpha} = \mathcal{F}_\alpha^{-1},\qquad \text{and} \qquad \mathcal{F}_{\alpha = \pi/2} = \mathcal{F}
\end{equation*}

As in the Fourier case, we calculate the canonical correlation Green's function in terms of $H$ and $G_0$, then write it in the form $V \Lambda V^{-1}$ with $\Lambda$ diagonal.
\begin{eqnarray*}
    \Sigma_{\chi}^{-1}\Sigma_{\chi|h} &=& (I + H G_0)^{-1} \\
    &=& (I + \eta \mathcal{F}^\dagger \tilde{G}_0^{-1/2}\mathcal{F}_\alpha \tilde{G}_0 \mathcal{F}_\alpha^\dagger \tilde{G}_0^{-1/2}\mathcal{F} \mathcal{F}^\dagger \tilde{G}_0 \mathcal{F})^{-1} \\ 
    &=& (I + \eta \mathcal{F}^\dagger \tilde{G}_0^{-1/2}\mathcal{F}_\alpha \tilde{G}_0 \mathcal{F}_\alpha^\dagger \tilde{G}_0^{1/2}\mathcal{F})^{-1} \\
    &=& (I + \eta (\mathcal{F}^\dagger \tilde{G}_0^{-1/2}\mathcal{F}_\alpha \tilde{G}_0^{1/2})\tilde{G}_0(\tilde{G}_0^{-1/2}\mathcal{F}_\alpha^\dagger \tilde{G}_0^{1/2}\mathcal{F}))^{-1} \\ 
    &=& (\tilde{G}_0^{-1/2}\mathcal{F}_\alpha^\dagger \tilde{G}_0^{1/2}\mathcal{F})^{-1}(I + \eta \tilde{G}_0)^{-1} (\mathcal{F}^\dagger \tilde{G}_0^{-1/2}\mathcal{F}_\alpha \tilde{G}_0^{1/2})^{-1} \\
    &=& V\Lambda V^{-1}
\end{eqnarray*}
\noindent
Hence we arrive at the eigendecomposition:
\begin{equation*}
    V(x,u) = [\mathcal{F}^\dagger \tilde{G}_0^{-1/2}\mathcal{F}_\alpha \tilde{G}_0^{1/2}](x,u)\,, \qquad \lambda(u) = \frac{1}{1 + \eta \tilde{G}_0(u)}
\end{equation*}

In the main text, we refrain from writing $V^\dagger$ as a kernel $V^\dagger(u,x)$, because it is discontinuous and divergent. This is more evident when it is expressed in integral form:
\begin{eqnarray*}
V^\dagger(u,x) &=& [\tilde{G}_0^{1/2}\mathcal{F}_\alpha^\dagger \tilde{G}_0^{-1/2}\mathcal{F}](u,x) \\ 
&=& (-2\pi i \sin\alpha)^{-d/2}(2\pi)^{-d/2}\sqrt{\frac{1}{1 + u^2}} \int \dd^dq\, \sqrt{1 + q^2} \exp\lrs{iq \cdot(u \csc\alpha - x) - \frac{i}{2}\cot(\alpha)(q^2 + u^2)}
\end{eqnarray*}

Where, e.g., $u^2 = u\cdot u = u_1^2 + u_2^2 +...+u_d^2$.
\end{widetext}


\bibliography{thebib.bib}

\end{document}